\documentclass[10pt,journal,compsoc]{IEEEtran}
\usepackage{makecell}
\usepackage{hyperref}
\usepackage{array}
\usepackage{graphicx,amssymb,amsmath}
\usepackage{multicol}
\usepackage[noadjust]{cite}
\usepackage{subfigure}
\usepackage{graphicx}
\usepackage{float}
\usepackage{url}
\usepackage{stfloats}
\usepackage{amsthm,pifont}
\usepackage{flushend}
\usepackage{cases,subeqnarray}
\usepackage{bm,multirow,bigstrut}
\usepackage{amsmath, amsthm, amssymb}
\usepackage{textcomp}
\usepackage{latexsym,bm}
\usepackage{setspace}
\usepackage{booktabs}
\usepackage{xcolor}
\usepackage{mathtools}
\usepackage{dsfont}
\usepackage{extarrows}
\usepackage{epsfig}
\usepackage{epsfig}
\usepackage{epstopdf}
\usepackage[noend]{algpseudocode}
\usepackage{algorithmicx,algorithm}
\usepackage{svg}
\usepackage[numbers,sort&compress]{natbib}
\theoremstyle{plain}

\theoremstyle{plain}
\newtheorem{rem}{Remark}

\usepackage{amsmath}

\usepackage{tablefootnote}
\usepackage{algorithmicx}
\IEEEoverridecommandlockouts

\begin{document}
	\title{{\color{black}Diffusion-based Reinforcement Learning for Edge-enabled AI-Generated Content Services}}
	\author{Hongyang~Du*, Zonghang Li*, Dusit~Niyato,~\IEEEmembership{Fellow,~IEEE}, Jiawen~Kang, Zehui Xiong, Huawei~Huang, and Shiwen~Mao,~\IEEEmembership{Fellow,~IEEE}
		\thanks{H.~Du, and D. Niyato are with the School of Computer Science and Engineering, the Energy Research Institute @ NTU, Interdisciplinary Graduate Program, Nanyang Technological University, Singapore (e-mail: hongyang001@e.ntu.edu.sg, dniyato@ntu.edu.sg). Z. Li is with the School of Information and Communication Engineering, University of Electronic Sciences and Technology of China, Chengdu, China (email: lizhuestc@gmail.com). J. Kang is with the School of Automation, Guangdong University of Technology, China (e-mail: kavinkang@gdut.edu.cn). Z. Xiong is with the Pillar of Information Systems Technology and Design, Singapore University of Technology and Design, Singapore (e-mail: zehui\_xiong@sutd.edu.sg). H. Huang is with the School of Software Engineering, Sun Yat-Sen University, Zhuhai, China (e-mail: huanghw28@mail.sysu.edu.cn). S. Mao is with the Department of Electrical and Computer Engineering, Auburn University, Auburn, USA (e-mail: smao@ieee.org). H. Du and Z. Li have equal contributions.}
	}
	\maketitle
	\vspace{-1.5cm}
	\begin{abstract}
		As Metaverse emerges as the next-generation Internet paradigm, the ability to efficiently generate content is paramount. 
		{\color{black}AI-Generated Content (AIGC) emerges as a key solution, yet the resource-intensive nature of large Generative AI (GAI) models presents challenges.}
		To address this issue, we introduce an AIGC-as-a-Service (AaaS) architecture, which deploys AIGC models in wireless edge networks {\color{black} to ensure broad AIGC services accessibility for Metaverse users.}
		Nonetheless, an important aspect of providing personalized user experiences requires carefully selecting AIGC Service Providers (ASPs) capable of effectively executing user tasks, {\color{black}which is complicated by environmental uncertainty and variability.
			Addressing this gap in current research, we introduce the AI-Generated Optimal Decision (AGOD) algorithm, a diffusion model-based approach for generating the optimal ASP selection decisions.
			Integrating AGOD with Deep Reinforcement Learning (DRL), we develop the Deep Diffusion Soft Actor-Critic (D2SAC) algorithm, enhancing the efficiency and effectiveness of ASP selection.}
		Our comprehensive experiments demonstrate that D2SAC outperforms seven leading DRL algorithms. 
		Furthermore, the proposed AGOD algorithm has the potential for extension to various optimization problems in wireless networks, positioning it as a promising approach for future research on AIGC-driven services.
		The implementation of our proposed method is available at: \url{https://github.com/Lizonghang/AGOD}.
	\end{abstract}
	
	\IEEEpeerreviewmaketitle
	\section{Introduction}
	\IEEEPARstart{T}{he} Turing Test, a renowned evaluation benchmark, was proposed by Alan Turing in his seminal paper~\cite{turing2009computing} to assess the {\textit{intelligence}} of machines, i.e., their ability to mimic human thinking and {\textit{generate}} content that can interact with humans. 
	Since then, the ability of {\color{black}Artificial Intelligence (AI)} to create content has become a fundamental research goal, as it is believed to be a key enabler for an epoch-making intelligence society. 
	This ambitious vision aligns with the requirements of Metaverse~\cite{wang2022survey}. 
	As we move towards a more immersive and interactive future Internet, the ability to generate vast amounts of high-quality digital content, e.g., user-defined avatars, becomes increasingly significant.
	
	Fortunately, AI-Generated Content (AIGC) has emerged as a powerful force driving innovation. According to a study by PriceWaterhouseCoopers, AI can increase global GDP by 14\% or approximately \$15.7 trillion by 2030~\cite{anand2019s}. This highlights the transformative impact of AIGC in driving economic growth and spurring technology adoption. For example, ChatGPT, a chatbot developed by OpenAI, has achieved remarkable success in generating human-like text~\cite{aljanabi2023chatgpt}. 
	Furthermore, Stable Diffusion, a text-to-image Generative AI (GAI) model launched in 2022 by Stability AI, can generate images in seconds conditioned on text descriptions~\cite{ulhaq2022efficient}. 
	With these capabilities, AIGC techniques are rapidly becoming essential for content creation and delivery, which is considered the ``engine'' in powering Metaverse~\cite{du2023enabling,lin2023blockchain}.
	
	Despite the remarkable advances in AIGC techniques, several challenges are associated with deployment~\cite{du2023age}. 
	One of the most significant issues is the increasing cost of developing and deploying AIGC models in user devices, e.g., head-mounted displays. 
	AIGC models require large datasets and complex architectures to achieve state-of-the-art performance, leading to massive resource consumption and longer training times~\cite{harshvardhan2020comprehensive}. 
	Furthermore, these models require high-end hardware and specialized software for training and inference, making it difficult for individuals to access and utilize AIGC in Metaverse. As such, the high cost limits the widespread adoption of AIGC.
	
	Another major obstacle stems from the diversity of users~\cite{du2022attention}. The Metaverse is expected to accommodate many user types, including those with varying cultural backgrounds, languages, and preferences. AIGC models must therefore be capable of generating content that is tailored to the individual user and meets their unique needs and expectations. Achieving this level of customization is challenging, as it requires a deep understanding of user behavior and online task scheduling mechanisms. 
	In general, on the way to building a human-centric Metaverse with the AIGC technique, the following two goals exist:
	\begin{itemize}
		\item[{\textbf{G1)}}] {\textit{Make AIGC a Metaverse support technology accessible from any device, anywhere, at any time}}
		\item[{\textbf{G2)}}] {\textit{Provide human-centric AIGC services, maximizing Metaverse user utilities while meeting users needs}}
	\end{itemize}
	To achieve the first goal ({\textbf{G1}}), one promising approach is to adopt the ``everything-as-a-service'' paradigm. Specifically, instead of distributing the trained AIGC models to user devices, they can be deployed on network edge servers, enabling the realization of ``AIGC-as-a-Service'' (AaaS) through the wireless edge networks. When a user requires AIGC services, the user can upload the demand to the network edge server, which executes the task through the AIGC model and sends the results back to the user. This approach has several advantages, including reducing the computational burden on user devices and providing flexible and scalable AIGC services. Furthermore, with the rapid advancement of wireless communication and computing technologies, the Sixth Generation (6G) of wireless networks is emerging as the next frontier in mobile communication systems, which are expected to provide ultra-high data speeds, ultra-low latency, and ultra-high dependability, enabling real-time responses to user requests. As a result, the deployment of AaaS can provide an efficient and reliable solution for delivering AIGC services to users while also enabling the development of new applications and services.
	
	However, the adoption of the AaaS approach poses a significant challenge to the second goal ({\textbf{G2}}), which is to provide human-centric AIGC services that maximize the utilities for the users. The challenge stems from the fact that various AIGC models possess different capabilities and are suited to specific tasks. For example, some AIGC models generate human-like images, while others perform better in producing natural scenery. Users also exhibit varying interests and preferences, and servers display varying computation capacities. Consequently, it becomes imperative to select {\color{black}the best AIGC Service Provider (ASP)} for many users, considering their specific requirements, personality, the computing resources available on the edge servers, and the attributes of the deployed AIGC models. By utilizing an efficient scheduling algorithm, it is possible to optimize the benefits of AaaS services for the users, enhancing their immersive experience and augmenting their engagement with Metaverse~\cite{du2023enabling,ren2022efficient}.
	
	Thus, a well-designed ASP selection algorithm is essential to achieve the two goals of providing ubiquitous and human-centric AIGC services. However, the difficulty in mathematically modeling both user utilities and AIGC model capabilities poses a significant challenge. Deep Reinforcement Learning (DRL)-based methods are a promising solution, but may not be efficient due to their dependence on exploration-exploitation trade-offs and potential convergence to suboptimal policies~\cite{osband2016deep}. To overcome these limitations, we propose a novel diffusion model-based AI-Generated Optimal Decision (AGOD) algorithm~\cite{du2023beyond}. {\textit{Similar to the AIGC technique in which diffusion models generate content, we adapt diffusion models to generate optimal decisions.}} The contributions of this paper are summarized as follows:
	\begin{itemize}
		\item We propose an architecture for AaaS that deploys AIGC models in the edge network, providing ubiquitous AIGC functionality to users in Metaverse (For {\textbf{G1}}).
		\item We propose the AGOD algorithm, empowered by diffusion models, to generate optimal decisions in the face of environmental uncertainty and variability (For {\textbf{G2}}).
		\item We apply our proposed AGOD to DRL, specifically in the form of the Deep Diffusion Soft Actor-Critic (D2SAC) algorithm, which achieves efficient and optimal ASP selection, thereby maximizing the user's subjective experience  (For {\textbf{G1}} and {\textbf{G2}}).
		\item We demonstrate the effectiveness of the proposed algorithm through extensive experiments, showing that D2SAC outperforms seven representative DRL algorithms, i.e., Deep Q-Network (DQN)~\cite{mnih2015human}, Deep Recurrent Q-Network (DRQN)~\cite{hausknecht2015deep}, Prioritized-DQN~\cite{schaul2015prioritized}, Rainbow~\cite{hessel2018rainbow}, REINFORCE~\cite{williams1992simple}, Proximal Policy Optimization (PPO)~\cite{schulman2017proximal}, and Soft Actor-Critic (SAC)~\cite{haarnoja2018soft} algorithms, not only in the studied ASP selection problem but also in various standard control tasks.
	\end{itemize}
	The rest of the paper is structured as follows: Section~\ref{rela} {\color{black}reviews} the related work. In Section~\ref{Saaas}, we introduce the AaaS concept and formulate the ASP selection problem. In Section~\ref{Sagod}, we propose the diffusion model-based AGOD algorithm. Section~\ref{Sdrl} presents the novel deep diffusion reinforcement learning algorithm, e.g., D2SAC, by applying AGOD in DRL. We conduct a comprehensive evaluation of the proposed D2SAC in Section~\ref{Sexp}. Finally, Section~\ref{Scon} concludes this paper.

	\section{Related Work}\label{rela}
	In this section, we provide a brief review of the related work, i.e., AIGC in Metaverse, diffusion model in optimization, and DRL.
	\subsection{Artificial Intelligence-Generated Content in Metaverse}
	Metaverse has gained significant attention as a future Internet. However, creating digital content is a prerequisite for establishing a symbiotic Internet between the virtual and real worlds. 
	Fortunately, AIGC technologies provide technical support for the rapid creation of digital content by leveraging the power of AI to automate the information creation process~\cite{du2023exploring}. This innovative content generation method represents a paradigm shift from traditional User-Generated Content (UGC) and Professionally Generated Content (PGC). Recent research has explored the potential of AIGC in empowering Metaverse. For example, to promote the construction of a virtual transportation system, the authors in~\cite{lin2023blockchain} propose a blockchain-aided semantic communication framework for AIGC services to facilitate interactions of the physical and virtual domains among virtual service providers and edge devices. Moreover, the authors in~\cite{liu2023block} present a blockchain-empowered framework to manage the life-cycle of edge AIGC products.
	Despite the significant potential of AIGC, the issue of enabling widespread access to huge AIGC models still needs to be solved~\cite{yue2022efficient}.

	\subsection{Diffusion Model in Optimization}
	Diffusion models, recognized as potent deep generative models, have become increasingly popular in machine learning, particularly in the image and video generation and molecule design~\cite{du2023beyond,ulhaq2022efficient}. These models aim to learn a given dataset's latent structure by modeling how data points diffuse through the latent space. In computer vision, neural networks have been trained to denoise images blurred with Gaussian noise by learning to reverse the diffusion process~\cite{ulhaq2022efficient}. 
	A groundbreaking approach called Diffusion Q-Learning (DQL) was introduced recently, using a conditional diffusion model to perform behavior cloning and policy regularization~\cite{wang2022diffusion}. The authors demonstrate the superior performance of their method compared to prior works in a 2D bandit example with a multi-modal behavior policy. However, it should be noted that DQL can only be used in offline DRL tasks with imitation learning. This limitation makes obtaining open datasets for online communication scheduling tasks impractical.
	More recently, a novel AI-generated incentive mechanism algorithm was proposed by authors in \cite{aigenerated2023du} to solve the utility maximization problem by generating optimal contract designs. The proposed diffusion model-based algorithm has been shown to outperform two deep reinforcement learning algorithms, i.e., PPO and SAC. However, both methods in \cite{wang2022diffusion,aigenerated2023du} are designed for continuous action space problems and cannot be applied in environments with discrete action spaces.
	
	\subsection{Deep Reinforcement Learning}
	{\color{black} DRL, an extension of Reinforcement Learning utilizing deep neural networks, excels at capturing state space representations. This capability empowers DRL agents to address complex and high-dimensional challenges, making it particularly effective for sequential decision-making problems~\cite{chen2018reinforcement}. The ASP selection problem, characterized by its online nature, presents a scenario where DRL's adaptability is particularly advantageous. DRL's dynamic learning framework allows it to efficiently adjust to unforeseen tasks that may emerge during operational processes, making it a highly suitable approach for the ASP selection challenge.}
	However, there are limitations to this method that can impede its effectiveness. In particular, the high computational requirements of DRL algorithms can be a challenge, especially for problems with large state or action spaces~\cite{arulkumaran2017deep}. In this case, the policy function in the DRL algorithm may not output optimal action decisions based on the current state. Therefore, an innovative approach is to incorporate the AIGC technique in generating optimal action decisions. 
	
	Building upon the limitations of existing research, we introduce an innovative solution to the ASP selection problem in the form of an AaaS approach. To this end, we leverage the power of the diffusion model and present the AGOD algorithm, which we then apply to DRL to propose the D2SAC algorithm.
	
	\section{AIGC Services in Wireless Networks}\label{Saaas}
	In this section, we introduce the AaaS in wireless edge networks, followed by the ASP selection problem formulation. Then, we introduce the human-aware utility function.
	
	\subsection{AIGC-as-a-Service}\label{aaasc}
	\begin{figure*}[t]
		\centering
		\includegraphics[width=0.9\textwidth]{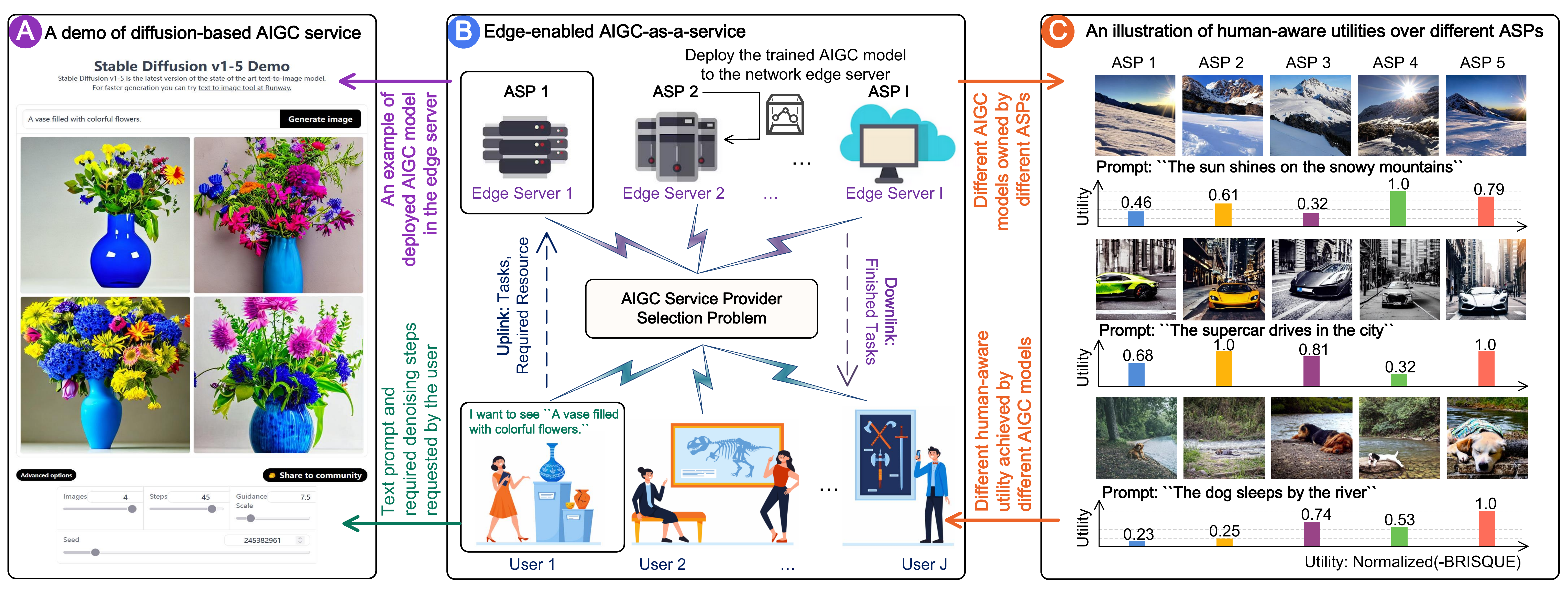}%
		\caption{{\color{black} The architecture of AIGC-as-a-Service in wireless edge networks. {\textbf{Part A}} is demo of AIGC service based on Stable Diffusion v1.5 as an example of deployable AIGC model for edge servers; {\textbf{Part B}} is network architecture of ASPs employing edge servers to deploy AIGC models for providing AaaS to users; {\textbf{Part C}} shows variation in user experience demonstrated by different outputs from the same text prompt on various AIGC models, highlighting the importance of ASP selection.}}
		\label{aaas}
	\end{figure*}
	AIGC techniques provide a fast and efficient content generation ability while reducing network resource consumption. AIGC models can help repair corrupted images, generate natural and realistic Augmented Reality/Virtual Reality/High-Definition (AR/VR/HD) video content for Metaverse users, and simplify the design of wireless transmission protocols. However, deploying AIGC models is typically challenging due to their large size and difficulty in training and deployment. To make AIGC services accessible from any device, anywhere, at any time, we propose deploying the AIGC model on network edge devices, as illustrated in Fig.~\ref{aaas} (Part B), to support AaaS. For instance, a Metaverse user can upload a generation request via the mobile device to an edge server. Then, the server sends the AIGC computation results after completing the task. Moreover, users can customize the computational resources required for their tasks when uploading them to the ASP. One example is given in Fig.~\ref{aaas} (Part A), the user interface of the stable diffusion model of the Hugging Face platform\footnote{The URL for Stable Diffusion v1-5 Demo in Hugging Face is {\url{https://huggingface.co/spaces/runwayml/stable-diffusion-v1-5}}.} allows users to specify the number of denoising steps for the diffusion model. Thus, the AaaS approach provides a scalable and efficient solution for wireless users to access AIGC services on demand, even on resource-constrained devices.
	However, to deploy AaaS, the following challenges still need to be addressed:
	\begin{itemize}
		\item[{\textbf{C1)}}] Users may access the AIGC service at their discretion and request customized computational resources, such as denoising steps of the diffusion model-based AIGC.
		\item[{\textbf{C2)}}] Performance evaluation of AIGC tasks is human-subjective and difficult to model mathematically.
		\item[{\textbf{C3)}}] The capacities of AIGC models deployed on network edge servers vary, as do the qualities of AIGC services offered by different ASPs and the computational resources available for each server, i.e., the maximum number of AIGC tasks that can be processed in a given time window.
	\end{itemize}
	Therefore, to improve the QoS of the entire AaaS system, an efficient and feasible algorithm for selecting an appropriate ASP is necessary. A high-quality AaaS system produces satisfactory results and reduces the likelihood of encountering problems or errors that could negatively impact the wireless network's performance. By selecting the optimal ASP, users can benefit from high-quality content generation services and fully leverage the potential of the wireless network with minimal errors and resource consumption.
	
	\subsection{AIGC Service Provider Selection}
	The ASP selection problem is analogous to a resource-constrained task assignment problem, where the aim is to allocate incoming tasks to available resources, satisfying resource constraints and maximizing overall utility. This problem is frequently encountered in wireless networks, where resources are scarce and their efficient utilization is crucial to achieving the desired performance, including task scheduling and resource allocation in wireless networks~\cite{sun2020dynamic,sun2022game,dai2022psaccf}.
	
\begin{figure}
		\centering
		\includegraphics[width=.42\textwidth]{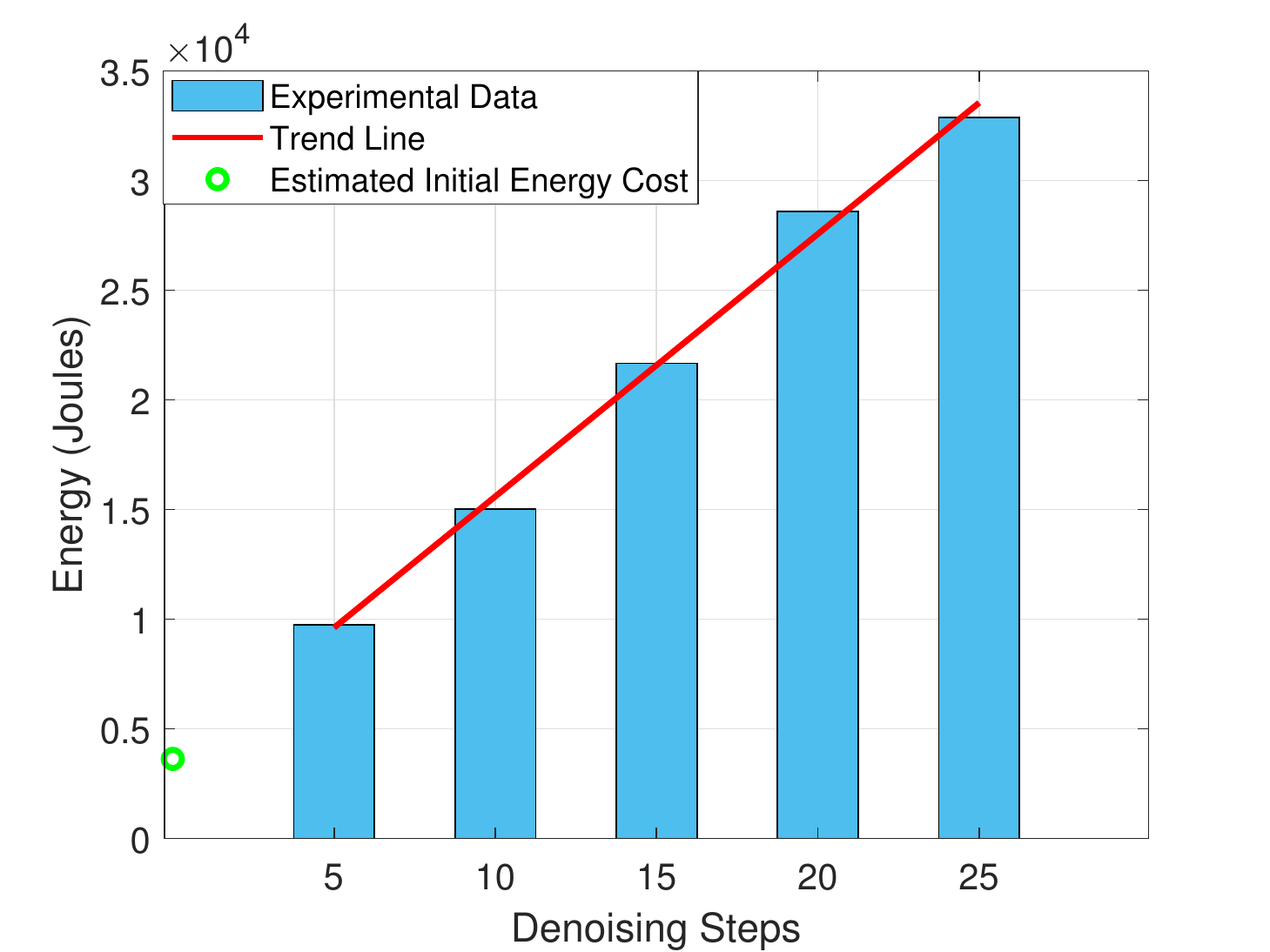}
		\caption{{\color{black} Energy cost versus diffusion steps for stable-diffusion-v1-4 model inference.}}
		\label{fig:energy}
	\end{figure}

	For the ASP selection, which can be framed as a resource-constrained task assignment problem, a set of sequential tasks $\mathcal{J}=\{j_1, j_2,\ldots,j_J\}$, a set of available ASPs $\mathcal{I}=\{ i_1, i_2,\ldots,i_I \}$, and the unique utility function $u_{i}(\cdot)$ of the $i^{\rm th}$ ASP are given. The objective is to find an assignment of tasks to ASPs, i.e., $\mathcal{A}=\left\{ a_1,\ldots,a_j,\ldots,a_J \right\}$, such that the overall users' utility $\mathcal{U}=\sum_{j=1}^J{u_{i}\left( T_j \right)}$ is maximized. Note that the utility $u_{i}\left( T_j\right) $ of the $j^{\rm th}$ task assigned to the $i^{\rm th}$ ASP is a function of the required resource $T_j$. 
	{\color{black}Without loss of generality, we consider that $T_j$ is the number of denoising steps of the diffusion model, which is positively correlated to the energy cost. 
		The reason is that each step of the diffusion model has energy consumption as it involves running a neural network to remove Gaussian noise~\cite{ho2020denoising}. 
		To empirically validate this relationship, we conducted experiments using a Dell Precision 5820 Tower equipped with an Intel Xeon W-2235 CPU. 
		Power metrics were meticulously recorded via HWiNFO64\footnote{HWiNFO64 (\url{https://www.hwinfo.com/}) is a hardware analysis and monitoring tool for Windows, presenting real-time information including fan speeds, voltages, power consumption, etc.} during the inference process of stable-diffusion-v1-4 model~\cite{stabdiff}. The results, illustrated in Fig.~\ref{fig:energy}, confirm a consistent increase in energy cost corresponding to the number of denoising steps, alongside an initial energy expenditure likely due to model initialization.}

	Furthermore, the utility function is human-aware, which is discussed in Section~\ref{saf}. The total availability of resources $\mathcal{T}_i$ $\left(i = 1,\ldots,I\right)$ for each ASP is considered.
	{\color{black} Note that, for illustrative purposes, we consider image-based AIGC that utilizes the diffusion model. However, our research approach is generalizable to other types of AIGC services, including those based on natural language processing (e.g., ChatGPT). One can substitute the relevant resources to be scheduled (e.g., GPU resources) and corresponding user utility functions as appropriate.}
	Mathematically, the ASP selection problem can be formulated as an integer programming problem with decision variables $a_j \left( \forall j\in\mathcal{J}\right) $ in $\mathcal{A}$, which represents the time series of task assignments to available ASPs. Additionally, $\hat{\mathcal{J}}_i$ denotes the set of running tasks on the $i^{\rm th}$ ASP at the time of assigning the current task. Thus, the problem can be formulated as
	\begin{align}\label{eq:utility-target}
		\underset{\mathcal{A}}{\max}\quad &\mathcal{U}=\sum_{j\in \mathcal{J}}{u_i\left( T_j \right)},\\
		\mathrm{s}.\mathrm{t}.\quad &i=a_j,\\
		&T_j+\sum_{j\prime\in \hat{\mathcal{J}}_i}{T_{j\prime}} \leqslant \mathcal{T}_i\quad \left( \forall i\in \mathcal{I} \right) , \label{eq:system-model-resource-constraint}\\
		& i = 1, \ldots I, \:{\text{and}} \:\: j = 1, \ldots J.
	\end{align}
	In this formulation, the resource constraints are incorporated through the constraint \eqref{eq:system-model-resource-constraint}, which specifies the limitations on the available resources. Note that failing to satisfy the constraint \eqref{eq:system-model-resource-constraint} can result in the crash of $i^{\rm th}$ ASP, causing the termination and restart of its running tasks.
	\begin{rem}
		The resource-constrained task assignment problem, i.e., \eqref{eq:utility-target}, is a well-known NP-complete problem \cite{mazzola1986resource}, which implies that finding an optimal solution in polynomial time is computationally infeasible. Moreover, the user can access the AaaS at their discretion, and the user utility is human-aware without mathematical expressions. Traditional mathematical methods are difficult to be applied. To address this challenge, different techniques, including greedy algorithms, genetic algorithms, and (meta-)heuristics algorithms, have been proposed to find an approximate solution. However, these techniques often assume that all tasks and their corresponding utility values are known in advance~\cite{desale2015heuristic}, which is impractical in ASP selection, where tasks arrive dynamically and in real time.
	\end{rem}
	
	In this case, the AaaS system scheduler must make real-time decisions while considering the current system state and the arrival of new tasks. Balancing the task assignments to available servers and reserving resources for future tasks is essential. Moreover, characteristics such as the utility value depend not only on the human-aware tasks but also on the assigned ASP's ability, which is unknown at the time of scheduling, making the problem more challenging than the online resource-constrained task assignment \cite{mehta2013online}.
	
	\subsection{Human-aware Utility Function}\label{saf}
	The utility value of a Metaverse user task is not known in advance. Instead, it is determined by considering human-aware content quality assessment techniques to the AIGC. Let us denote $\mathcal{F}_i(T_j)$ as the forward function of the AIGC model of the $i^{\rm th}$ ASP and $\mathcal{G}(\cdot)$ as the human-aware content quality assessment function. Then, the utility value $u_i(T_j)$ of the $j^{\rm th}$ task on the $i^{\rm th}$ ASP can be expressed as
	\begin{equation}\label{eq:utility-func}
		u_i(T_j)=\mathcal{G}(\mathcal{F}_i(T_j)), \:\: \left( i = 1, \ldots I, \:{\text{and}} \:\: j = 1, \ldots J\right).
	\end{equation}
	Taking the image-based AIGC service as an example, the AI model can generate images according to the text prompt uploaded by users or impair the distorted images. Without the loss of generality, the human-aware content quality assessment function $\mathcal{G}(\cdot)$ could be the Blind/Referenceless Image Spatial Quality Evaluator (BRISQUE), which is designed to be human-aware with aims to predict the image quality based on how humans perceive image quality. BRISQUE is trained on a dataset of natural images perceived as high quality by human observers, which can extract features relevant to human perception of image quality, such as contrast, sharpness, and texture. Therefore, BRISQUE is considered a no-reference (or blind) image quality assessment model that does not require a reference image to compare against. This makes BRISQUE more practical for real-world applications where a reference image may not be available or practical to use as the reference. By being human-aware, BRISQUE provides a reliable and objective measure of image quality.
	
	An illustration of the utility distribution among different ASPs in our case is presented in Fig.~\ref{aaas} (Part C). We can observe that there is a significant variance in human-aware utility values between ASPs, highlighting the importance of users selecting a well-suited ASP.
	
	\section{AI-Generated Optimal Decision}\label{Sagod}
	In this section, we propose the AGOD algorithm that generates optimal discrete decisions starting from Gaussian noise with the help of the diffusion model.
	\subsection{Motivation of AGOD}
	{\color{black} The discrete variables in the ASP selection problem present a unique challenge: the solution set is finite and discrete, making traditional optimization techniques for continuous variables ineffective~\cite{arulkumaran2017deep}. In this scenario, unlike the gradual progression toward optimality offered by continuous variables, discrete variables necessitate jumping from one distinct solution to another. This characteristic turns the problem into a combinatorial one, where the solution space grows exponentially with each added variable, rendering exhaustive searches impractical for large-scale problems~\cite{chen2018reinforcement,arulkumaran2017deep}. Resorting to continuous optimization by ignoring the discrete nature of decision variables only yields inaccurate and suboptimal results. This necessitates the development of novel optimization techniques adept at handling discrete variables and the complexity of combinatorial optimization, outperforming existing DRL algorithms in navigating this intricate and expansive solution space.}

	{\color{black}The Denoising Diffusion Probabilistic Model (DDPM), a framework originally for image generation, inspires our approach to optimize discrete decision solutions~\cite{du2023beyond}.
		It involves gradually adding noise to the data until the data is entirely Gaussian noise (the forward process). Then, the model learns to reverse the diffusion process to recover the original image (the reverse process). 
		Motivated by DDPM's exceptional generative capabilities, we aim to develop a diffusion-based optimizer for generating discrete decision solutions.
		The diffusion model's inherent ability to incorporate conditioning information into the denoising process enhances its applicability and precision~\cite{du2023beyond}. More importantly, the potential interaction between the diffusion model and DRL represents a complementary and mutually enhancing relationship, allowing both methods to benefit, thereby broadening the effectiveness of discrete decision optimization in complex and dynamic environments.
	}
	
\begin{figure*}
		\centering
		\includegraphics[width=.9\textwidth]{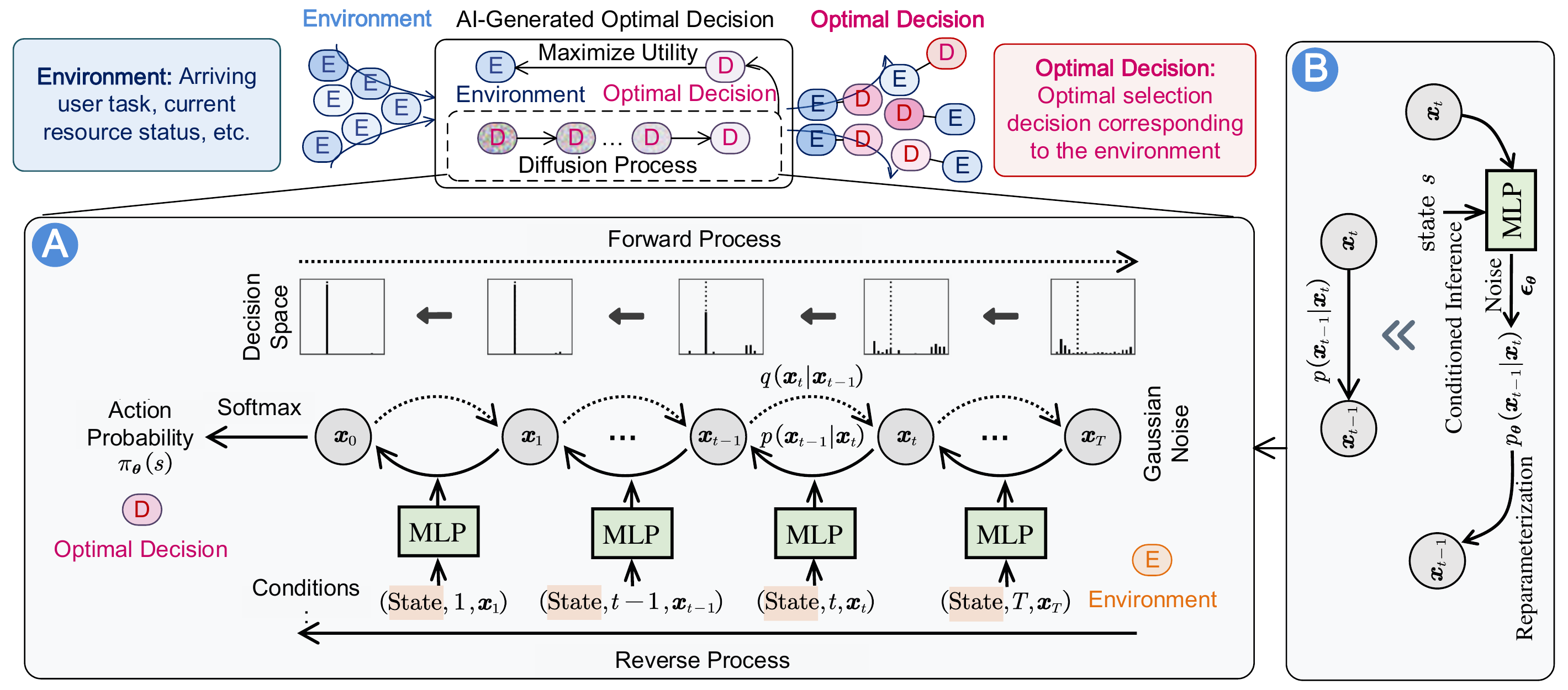}
		\caption{{\color{black}Illustration of the AGOD algorithm with the conditioned diffusion process.}}
		\label{fig:diffusion-process}
	\end{figure*}

	In the decision-making problem, the decision scheme can be expressed as a set of discrete probabilities for choosing each decision. The constraints and task-related factors affecting the optimal decision scheme can be considered the {\textit{environment}}. According to the diffusion model, an optimal decision solution under the current environment can keep increasing the noise until it becomes Gaussian, known as the forward process of probability noising~\cite{du2023beyond}. 
	Then, in the reverse process of probability inference, the {\textit{optimal decision generation network}}, i.e., $\pi _{\boldsymbol{\theta }}(\cdot)$, can be viewed as a denoiser that starts with Gaussian noise and recovers the optimal decision solution, i.e., $\boldsymbol{x}_0$, based on the environment condition, i.e., $s$. An illustration of the diffusion process is provided in Fig.~\ref{fig:diffusion-process}. 
	In the following, we present the forward process and propose the AGOD algorithm as the reverse process.
	
	\subsection{The Forward Process of Probability Noising}
	As the decision scheme output $\boldsymbol{x}_0=\pi _{\boldsymbol{\theta }}\left( s\right) \sim \mathbb{R}^{|\mathcal{A}|}$ is the probability distribution of each decision being selected under the observed environment state $s$, we represent the discrete vector of the distribution at step $t$ in the forward process as $\boldsymbol{x}_t$, which have the same dimensionality as $\boldsymbol{x}_0$. 
	Given a target probability distribution $\boldsymbol{x}_0$, the forward process adds a sequence of Gaussian noises at each step to obtain $\boldsymbol{x}_1,\boldsymbol{x}_2,\ldots,\boldsymbol{x}_T$. {\color{black} The transition from $\boldsymbol{x}_{t-1}$ to $\boldsymbol{x}_t$ is defined as a normal distribution with mean $\sqrt{1-\beta_t}\boldsymbol{x}_{t-1}$ and variance $\beta_t\mathbf{I}$ as~\cite{ho2020denoising}}
	{\color{black} \begin{equation}
			q\left( \boldsymbol{x}_t|\boldsymbol{x}_{t-1} \right) =\mathcal{N}\left( \boldsymbol{x}_t;\sqrt{1-\beta _t}\boldsymbol{x}_{t-1},\beta _t\mathbf{I} \right),
		\end{equation}
		where $t = 1,\ldots, T$,} $\beta _t=1-e^{-\frac{\beta _{\min}}{T}-\frac{2t-1}{2T^2}\left( \beta _{\max}-\beta _{\min} \right)}$ represents the forward process variance controlled by the Variational Posterior (VP) scheduler~\cite{ho2020denoising}. 
	
	As $\boldsymbol{x}_t$ depends only on $\boldsymbol{x}_{t-1}$ at the previous step, the forward process can be  considered a Markov process, and {\color{black} the distribution $\boldsymbol{x}_{T}$ given $\boldsymbol{x}_0$ can be formed as the product of transitions $q\left( \boldsymbol{x}_t|\boldsymbol{x}_{t-1} \right)$ over {\color{black} denoising step} as~\cite{ho2020denoising}}
	\begin{equation}
		q\left( \boldsymbol{x}_{T}|\boldsymbol{x}_0 \right) =\prod_{t=1}^T{q\left( \boldsymbol{x}_t|\boldsymbol{x}_{t-1} \right)}.
	\end{equation}
	
	The forward process is not actually executed, but it establishes the mathematical relationship between $\boldsymbol{x}_0$ and any $\boldsymbol{x}_t$ as
	\begin{equation}
		\label{eq:relationship-between-xt}
		\boldsymbol{x}_t=\sqrt{\bar{\alpha}_t}\boldsymbol{x}_0+\sqrt{1-\bar{\alpha}_t}\boldsymbol{\epsilon },
	\end{equation}
	where $\alpha_t=1-\beta_t$, $\bar{\alpha}_t=\prod_{k=1}^t{\alpha_k}$ is the cumulative product of $\alpha_k$ over previous {\color{black} denoising step} $k$ $(\forall k\le t)$, and $\boldsymbol{\epsilon }\sim \mathcal{N}\left( \mathbf{0},\mathbf{I} \right)$ is a standard normal noise. As $t$ increases, $\boldsymbol{x}_T$ becomes purely noise with a normal distribution of $\mathcal{N}(\mathbf{0},\mathbf{I})$. However, note that optimization problems in wireless network often lack a dataset of optimal decision solutions, i.e., $\boldsymbol{x}_0$, to be used for the forward process. Consequently, the forward process is not performed in AGOD.
	
	\subsection{The Reverse Process of Probability Inference}
	The reverse process, also called the sampling process, aims to infer the target $\boldsymbol{x}_0$ from a noise sample $\boldsymbol{x}_T\sim\mathcal{N}(\mathbf{0},\mathbf{I})$ by removing noise from it. In our AGOD algorithm, the purpose is to infer the optimal decision scheme from the noise sample.
	The transition from $\boldsymbol{x}_t$ to $\boldsymbol{x}_{t-1}$ is defined as $p\left( \boldsymbol{x}_{t-1}|\boldsymbol{x}_t \right)$, which cannot be calculated directly. However, it follows a Gaussian distribution as given by
	\begin{equation}
		\label{eq:p-transition}
		p_{\boldsymbol{\theta }}\left( \boldsymbol{x}_{t-1}|\boldsymbol{x}_t \right) =\mathcal{N}\left( \boldsymbol{x}_{t-1};\boldsymbol{\mu }_{\boldsymbol{\theta }}\left( \boldsymbol{x}_t,t,s \right) ,\tilde{\beta}_t\mathbf{I} \right), 
	\end{equation}
	where the mean $\boldsymbol{\mu }_{\boldsymbol{\theta }}$ can be learned by a deep model, and $\tilde{\beta}_t=\frac{1-\bar{\alpha}_{t-1}}{1-\bar{\alpha}_t}\beta _t$ is a deterministic variance amplitude that can be easily calculated~\cite{ho2020denoising}. 
	
	By applying the Bayesian formula, we transform the calculation of the reverse process into the calculation of the forward process and reformat it into the form of a Gaussian probability density function. Then, we obtain the mean as follows,
	{\color{black} \begin{align}
			\label{eq:mu-direct-form}
			\boldsymbol{\mu }_{\boldsymbol{\theta }}\left( \boldsymbol{x}_t,t,s \right) =\frac{\sqrt{\alpha _t}\left( 1-\bar{\alpha}_{t-1} \right)}{1-\bar{\alpha}_t}\boldsymbol{x}_t+\frac{\sqrt{\bar{\alpha}_{t-1}}\beta _t}{1-\bar{\alpha}_t}\boldsymbol{x}_0,
		\end{align}
		where $t = 1,\ldots,T$.} According to \eqref{eq:relationship-between-xt}, the reconstructed sample $\boldsymbol{x}_0$ can be directly obtained via
	\begin{equation}
		\label{eq:x0}
		\boldsymbol{x}_0=\frac{1}{\sqrt{\bar{\alpha}_t}}\boldsymbol{x}_t-\sqrt{\frac{1}{\bar{\alpha}_t}-1}\cdot \tanh \left( \boldsymbol{\epsilon }_{\boldsymbol{\theta }}(\boldsymbol{x}_t,t,s) \right),
	\end{equation}
	where $\boldsymbol{\epsilon}_{\boldsymbol{\theta}}(\boldsymbol{x}_t,t,s)$ is a deep model parameterized by $\boldsymbol{\theta }$ that generates denoising noises conditioned on the observation $s$. The generated noise is scaled to be small through the application of the hyperbolic tangent activation, as it may result in a high level of noise in $\boldsymbol{x}_0$, making it difficult to identify the true underlying action probability.
	
	The reverse process introduces a new source of noise $\boldsymbol{\epsilon}_{\boldsymbol{\theta}}$ at each {\color{black} denoising step} $t$, and they are independent of the noise $\boldsymbol{\epsilon}$ added in the forward process. This makes us unable to calculate $\boldsymbol{x}_0$ by directly using \eqref{eq:x0}. Instead, we apply \eqref{eq:x0} into \eqref{eq:mu-direct-form} to estimate the mean
	\begin{align}
		\label{eq:reverse-mean}
		\boldsymbol{\mu }_{\boldsymbol{\theta }}\left( \boldsymbol{x}_t,t,s \right) =\frac{1}{\sqrt{\alpha _t}}\left( \boldsymbol{x}_t-\frac{\beta _t \tanh \left( \boldsymbol{\epsilon }_{\boldsymbol{\theta }}(\boldsymbol{x}_t,t,s) \right)}{\sqrt{1-\bar{\alpha}_t}} \right).
	\end{align}
	Then, we can sample $\boldsymbol{x}_{t-1}$ from the reverse transition distribution $p(\boldsymbol{x}_t)p_{\boldsymbol{\theta }}\left( \boldsymbol{x}_{t-1}|\boldsymbol{x}_t \right)$, and further use the cumulative product over $t=T,T-1,\ldots,1$ to obtain the generation distribution $p_{\boldsymbol{\theta }}\left( \boldsymbol{x}_0 \right)$ as follows,
	\begin{equation}
		p_{\boldsymbol{\theta }}\left( \boldsymbol{x}_0 \right) =p\left( \boldsymbol{x}_T \right) \prod_{t=1}^T{p_{\boldsymbol{\theta }}\left( \boldsymbol{x}_{t-1}|\boldsymbol{x}_t \right)},
	\end{equation}
	where $p\left( \boldsymbol{x}_T \right) $ is a standard Gaussian distribution. Once we have the generation distribution $p_{\boldsymbol{\theta }}\left( \boldsymbol{x}_0 \right)$, we can sample the output $\boldsymbol{x}_0$ from it.

	It is a common challenge in training generative models with stochasticity that gradients cannot be back-propagated through the random variable in the operation of sampling from a distribution. To address this issue, we employ reparameterization, which decouples the randomness from the distribution parameters. Specifically, the following update rule is used instead,
	\begin{equation}
		\label{eq:xt-update-rule}
		\boldsymbol{x}_{t-1}=\boldsymbol{\mu }_{\boldsymbol{\theta }}\left( \boldsymbol{x}_t,t,s \right) +\left( \tilde{\beta}_t/2 \right) ^2\odot \boldsymbol{\epsilon }, 
	\end{equation}
	where $\boldsymbol{\epsilon }\sim \mathcal{N}\left( \mathbf{0},\mathbf{I} \right) $. By iteratively applying the reverse update rule, i.e., \eqref{eq:xt-update-rule}, we can obtain all $\boldsymbol{x}_t$ $\left( \forall t, 0\le t<T\right)$, and in particular, the output sample $\boldsymbol{x}_0$, from a randomly generated normal noise.
	
	Finally, we apply the softmax function to $\boldsymbol{x}_0 $ to convert it into a probability distribution as
	\begin{equation}
		\label{eq:softmax-output}
		\pi _{\boldsymbol{\theta }}\left( s \right) =\left\{\frac{e^{\boldsymbol{x}_{0}^{i}}}{\sum_{k=1}^\mathcal{A}{e^{\boldsymbol{x}_{0}^{k}}}},\forall i\in \mathcal{A} \right\}.
	\end{equation}
	The elements in $\pi _{\boldsymbol{\theta }}\left( s \right)$ correspond to the probability of selecting each action.

	When implementing AGOD in practical systems, we first compute the mean $\boldsymbol{\mu }_{\boldsymbol{\theta }}$ of the reverse transition distribution $p_{\boldsymbol{\theta }}\left( \boldsymbol{x}_{t-1}|\boldsymbol{x}_t \right)$, as defined in \eqref{eq:p-transition} and \eqref{eq:reverse-mean}, and then obtain the distribution $\boldsymbol{x}_{t-1}$ using the update rule in \eqref{eq:xt-update-rule}. Next, we can derive the probability distribution of the optimal decision $\boldsymbol{x}_0$ using \eqref{eq:softmax-output}. However, in DDPM, the optimization objective is the Mean Squared Error (MSE) loss, which requires labeled images as targets \cite{ho2020denoising}. This requirement poses significant challenges in real decision-making problems in wireless networks. Therefore, AGOD needs to learn in an exploratory manner, and the training goal of the denoising network shifts from minimizing the error with labeled data to maximizing the value of the decision scheme, i.e., being able to maximize the optimization objective. 
	One possible approach proposed by authors in~\cite{aigenerated2023du} is to construct a {\textit{decision value network}} whose output assesses the utility resulting from the decision scheme, i.e., the output of the {\textit{optimal decision generation network}}. Then, the two networks can be trained jointly. {\color{black} However, the approach in~\cite{aigenerated2023du} is for the case when the decision valuables are continuous valuables.} 
	
	{\color{black}Leveraging AGOD's adaptability, we aim to enhance the optimization potential by integrating AGOD into advanced DRL algorithms, specifically within the SAC framework. The SAC's efficiency and stable policy learning complement AGOD's generative strengths. This integration enriches the SAC model with AGOD's exploration and learning capabilities, leading to the development of D2SAC as a diffusion-based DRL algorithm.}

	\section{{\color{black} Diffusion-based Reinforcement Learning}}\label{Sdrl}
	In this section, we model the ASP selection problem and present our innovative deep diffusion reinforcement learning algorithm, D2SAC, by applying the AGOD in the action policy.
	
	\subsection{Problem Modeling}
	Recall that we have a series of tasks, $\mathcal{J}$, and a set of available ASPs, $\mathcal{I}$. The objective is to assign tasks to ASPs in a way that maximizes the overall utility, denoted as $\mathcal{U}$, where the utility of each task assigned to an ASP is a function of the required resource $T_j$. {\color{black}We consider resource limitations of each ASP, acknowledging that an ASP can only handle a finite number of tasks due to its resource constraints. Exceeding these resources risks ASP failure and the potential restart of tasks. This reality makes the {\textit{Markov Decision Process (MDP)}} framework particularly suitable for the ASP selection problem~\cite{chen2021deep}. MDP captures the sequential nature of decision-making and how each task assignment influences future rewards and actions. The unpredictable nature of task arrivals further justifies an MDP-based approach. This method enables real-time decision-making, considering the current system state and the need to allocate resources for future tasks, ensuring a balanced and sustainable task distribution among ASPs.}
	
	Given an initial state $s_0$, the agent transitions from one state $s_l \in \mathcal{S}$ to the next $s_{l + 1} \in \mathcal{S}$ at each step $l = 0, 1, \ldots, L$, by taking an action $a_l \in \mathcal{A}$ and receiving a reward $r_l \in \mathcal{R}$ in the environment. Here, the action decision is chosen according to the policy. We use the diffusion model in AGOD, i.e., $\pi_{\boldsymbol{\theta}}$, as the action policy.
	The aim is to maximize the accumulated reward, $R(s_0,\pi_{\boldsymbol{\theta}})$, defined as the expected sum of discounted rewards as
	\begin{equation}
		\label{eq:cumulative-reward}
		R\left( s_0,\pi _{\boldsymbol{\theta }} \right) =\mathbb{E}\left[ \sum_{l=0}^L{\gamma ^lr_l}|s_0,\pi _{\boldsymbol{\theta }} \right],
	\end{equation}
	where $\boldsymbol{\theta}$ are the parameters of the diffusion policy network, $\gamma \in [0,1]$ is the discount factor that determines the importance of future rewards relative to immediate rewards, $L$ is the number of transitions in an episode, and $\mathcal{P}$ is the transition probability of states.
	In this manner, the MDP model for our problem can be formally described as a tuple $(\mathcal{S}, \mathcal{A}, \mathcal{P}, \mathcal{R})$.
	
	\textbf{a) State Space.}
	The state space $\mathcal{S}$ in our problem contains the environment information to make the decision. The state of the agent $s\in\mathcal{S}$ is composed of two feature vectors, one representing the arriving user task, $s^{\mathrm{T}}$, and one representing the current resource status of all ASPs, $s^{\mathrm{A}}$.
	The feature vector $s^{\mathrm{T}}$ encodes the resources $T$, i.e., {\color{black} denoising step}, required by the task and its estimated completion time $o$, which is represented as $s^{\mathrm{T}}=[T,o]$.
	The feature vector $s^{\mathrm{A}}$ includes the total available resources $\mathcal{T}_i$ and the currently available resources $\tilde{\mathcal{T}}_i$ of each of the $I$ ASPs, which is defined as $s^{\mathrm{A}}=[{\mathcal{T}_i,\tilde{\mathcal{T}}_i}|\forall i\in\mathcal{I}]$.
	Finally, these two feature vectors are concatenated to form the state vector $s$ as $s=[s^{\mathrm{T}},s^{\mathrm{A}}]$.
	{\color{black} The values of} $T$, $o$, $\mathcal{T}_i$, and $\tilde{\mathcal{T}}_i$ are normalized to the range $(0, 1)$ before being fed into the AGOD network, i.e., policy network $\pi_{\boldsymbol{\theta}}(s)$, to ensure stable training.
	
	\textbf{b) Action Space.} 
	The action space $\mathcal{A}$ is defined as the set of all possible decisions that can be made by the agent. In the ASP selection problem, the action taken by the agent, $a\in\mathcal{A}$, represents the assignment of the current Metaverse user task to one of the $I$ available ASPs. Specifically, the action space is an integer space with values ranging from 1 to $I$. The action $a$ is determined by the AGOD network, i.e., $\boldsymbol{\pi}_{\boldsymbol{\theta}}(s)$, which generates a vector of $I$ elements with the current state $s$ as the input. Each element of the vector represents the probability of selecting a particular ASP, i.e., $a\sim\boldsymbol{\pi}_{\boldsymbol{\theta}}(s)$.
	Note that, during evaluation, the ASP with the highest probability is selected, i.e., 
	\begin{equation}
		a=\underset{i}{\mathrm{arg}\max}\left\{ \pi _{\boldsymbol{\theta }}^{i}(s),\forall i\in \mathcal{I} \right\},
	\end{equation}
	where $\pi _{\boldsymbol{\theta }}^{i}(s)$ represents the probability of selecting ASP $i$.
	
	\textbf{c) Reward Function.}
	{\color{black}The reward $r \in \mathcal{R}$ is a scalar representing the immediate benefit received upon executing action $a$ in state $s$. The reward function $r\left(s,a\right)$ comprises two parts: the AIGC quality reward $r^{\mathrm{R}}$ and the crash penalty $r^{\mathrm{P}}$. Specifically, $r^{\mathrm{R}}$ reflects the generated content's quality, determined using the content quality assessment methods detailed in \eqref{eq:utility-func}. 
		To discourage low-quality content, the utility value $u_i\left(T_j\right)$ is adjusted by a baseline score $\hat{r}^{\mathrm{R}}$ from a noise sample, resulting in $r^{\mathrm{R}} = u_i\left(T_j\right) - \hat{r}^{\mathrm{R}}$. 
		The crash penalty $r^{\mathrm{P}}$, imposed on actions that overload the ASP causing task interruptions, consists of a fixed penalty $\hat{r}^{\mathrm{P}}_{\mathrm{F}}$ and an additional penalty $\hat{r}^{\mathrm{P}}_{\mathrm{I}}$ proportional to the progress of ongoing tasks $\hat{\mathcal{J}}$ as}
	\begin{equation}
		r^{\mathrm{P}}=\hat{r}^{\mathrm{P}}_{\mathrm{F}} - \sum_{j^\prime\in\hat{\mathcal{J}}}\hat{r}^{\mathrm{P}}_{\mathrm{I}}(j^\prime).
	\end{equation}
	We set $\hat{r}^{\mathrm{P}}_{\mathrm{F}}=2$ by default and $\hat{r}^{\mathrm{P}}_{\mathrm{I}}(j')$ as the multiply of $\hat{r}^{\mathrm{P}}_{\mathrm{F}}$ and the remaining progress of task $j'$ when it was interrupted.
	Incorporating the fixed penalty value $\hat{r}^{\mathrm{P}}_{\mathrm{F}}$ discourages the agent from taking actions that may cause a crash. The additional penalty $\hat{r}^{\mathrm{P}}_{\mathrm{I}}(j^{\prime})$ is associated with the interrupted task $j^{\prime}$, serving as incentive for the agent to refrain from disrupting ongoing tasks. Together, these penalties help to promote system stability.
	Finally, the reward $r$ returned by the environment can be represented as the sum of the reward and penalty as $r = r^{\mathrm{R}} - r^{\mathrm{P}}$. 
	{\color{black}In Section~\ref{Sexp}, we differentiate between `training reward,' which affects the learning process and policy optimization during training, and `test reward,' which evaluates the learned policy's generalization and robustness in new environments.}
	
	\textbf{d) Transition Function.} The transition function, represented by $p(s'|s,a)\in\mathcal{P}$, defines the probability of transitioning from the current state $s$ to the next state $s'$ after taking action $a$. The state transition model is intricate and cannot be mathematically formulated in our scenario. Instead, it relies on the unpredictable variables inherent in practical wireless network environments. The arrival of novel and unfamiliar tasks, the allocation of tasks to ASPs, and the successful or failed execution of tasks all influence state transitions.
	
	\textbf{e) Initialization and Termination.} 
	Every observation originates from the initial state $s_0$, and the agent begins acting based on it. $s_0$ is set as $(T_0,o_0,\mathcal{T}_1,1,\mathcal{T}_2,1,\ldots,\mathcal{T}_I, 1)$, with $T_0$ representing the required resources and $o_0$ denoting the estimated completion time of the first task. The repeated $(\mathcal{T}_i,1)$ of $I$ ASPs indicates that no ongoing tasks exist.
	The environment progresses from one state to another based on the actions taken by the agent until a termination criterion is met. To facilitate the policy network training, we introduce a termination condition by specifying a maximum number of transitions $L$ for each episode.
	
	Based on the above definitions, we present the overall goal of our problem, which is to train the parameters $\boldsymbol{\theta}^*$ of the AGOD network that maximizes the expected cumulative reward defined in \eqref{eq:cumulative-reward} as
	\begin{equation}
		\label{eq:maximize-goal}
		\boldsymbol{\theta }^*=\mathrm{arg}\max_{\boldsymbol{\theta }} \,\,\mathbb{E}\left[ \sum_{l=0}^L{\gamma ^l\left( r_l+\alpha H(\pi _{\boldsymbol{\theta }}(s_l)) \right)} \right] ,
	\end{equation}
	where the expectation is taken over all initial states $s_0$, and $H(\pi_{\boldsymbol{\theta}}(s_l))$ is called the action entropy regularization~\cite{haarnoja2018soft}, with $\alpha$ known as the temperature. The $H(\pi_{\boldsymbol{\theta}}(s_l))$ encourages the agent to explore more diverse actions.
	To take advantage of the efficient parallel computing capabilities of GPUs, we reverse the goal \eqref{eq:maximize-goal} by transforming the maximization problem into a minimization problem as 
	\begin{equation}
		\label{eq:minimize-goal}
		\boldsymbol{\theta }^*=\mathrm{arg}\min_{\boldsymbol{\theta }} \,\,-\mathbb{E}\left[ \sum_{l=0}^L{\gamma ^l\left( r_l+\alpha H(\pi _{\boldsymbol{\theta }}(s_l)) \right)} \right] .
	\end{equation}
	In solving the goal \eqref{eq:minimize-goal}, the agent strives to balance the trade-off between achieving high utility of task assignment and avoiding crashes to ASPs. Thus, the agent continuously updates the AGOD network parameters $\boldsymbol{\theta}$ based on the experience it gains during training.
	
	\subsection{Algorithm Architecture}
\begin{figure}[t]
		\centering
		\includegraphics[width=0.47\textwidth]{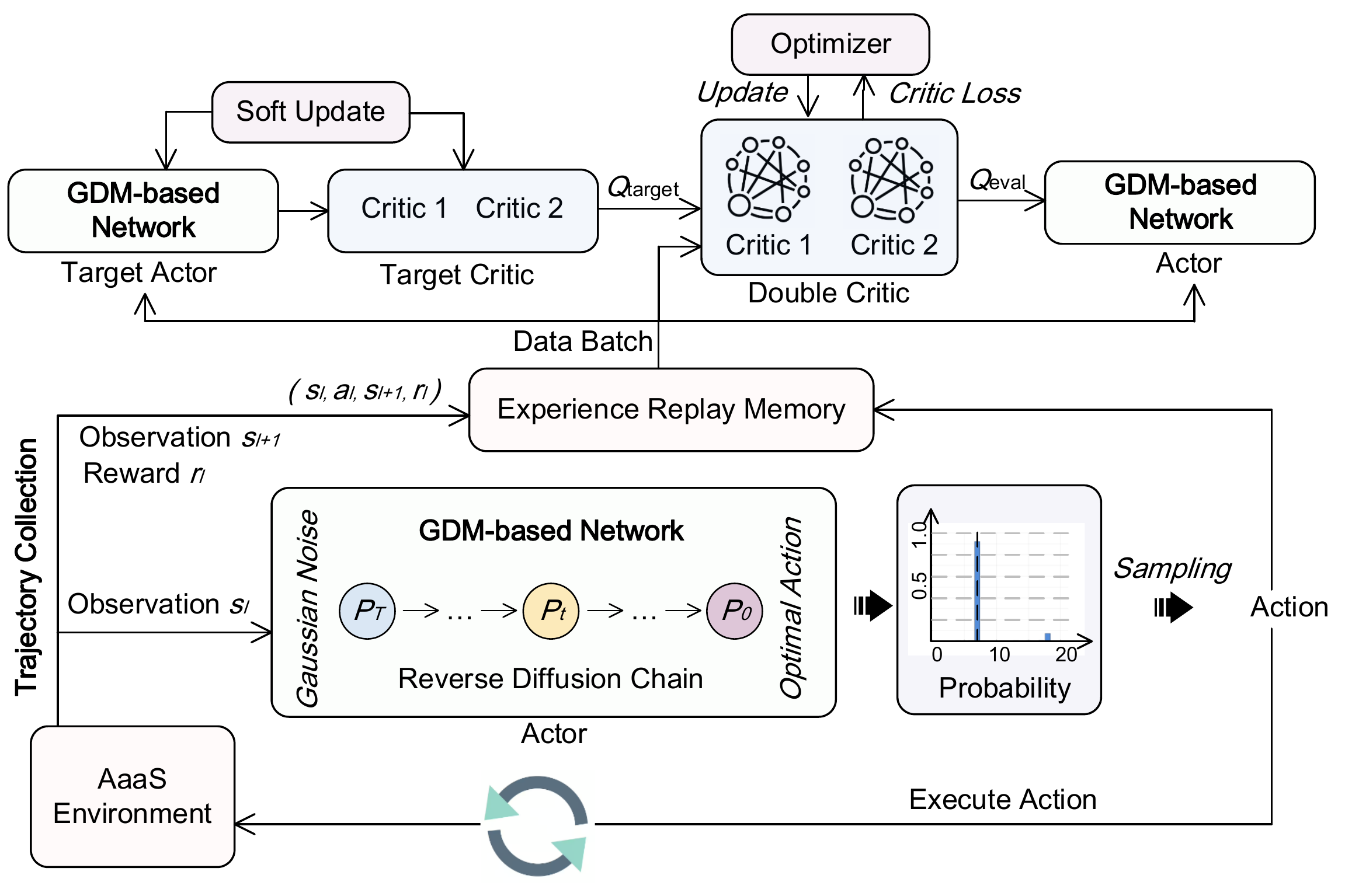}
		\caption{{\color{black} The overall architecture of the D2SAC~algorithm.}}
		\label{fig:alg-arch}
	\end{figure}

	The algorithm architecture of D2SAC, as shown in Fig.~\ref{fig:alg-arch}, consists of several components that work together to optimize the policy, i.e., an actor-network, a double critic network, a target actor, a target critic, an experience replay memory, and the environment.
	
	\textit{Trajectory Collection.}
	In this process, the agent starts by observing the environment and obtaining the initial observation $s_0$. The agent then collects $C$ transitions by iteratively generating and executing action decisions in the environment. These transitions are regarded as experiences and added to the experience replay memory $\mathcal{D}$, which has a capacity of $D=|\mathcal{D}|$. 
	More specifically, at each environment step $l$, the actor takes in the observation $s_l$ and outputs a discrete probability distribution ${\pi}_{\boldsymbol{\theta}}(s_l)$ over all possible actions $\mathcal{A}$. The agent then samples an action $a_l\sim{\pi}_{\boldsymbol{\theta}}(s_l)$ from this distribution and feeds it into the environment. The environment takes action, transits to state $s_{l+1}$, and returns an immediate reward $r_l$ as feedback to the agent. The agent records this experience with the transition tuple $(s_l,a_l,s_{l+1},r_{l})$ into the experience replay memory. These steps are repeated $C$ times before the policy improvement step.
	
	\textit{Diffusion Model-based AGOD as the Policy.}
	In D2SAC, the core of the actor-network ${\pi}_{\boldsymbol{\theta}}(s_l)$ is the diffusion model-based AGOD, rather than a conventional Multi-Layer Perception (MLP). AGOD enables effective representation encoding of the observation $s_l$, by utilizing $s_l$ as the input condition. This way, the diffusion process can effectively capture the dependencies between the observation and the action space.
	
	\textit{Experience Replay Memory.}
	Experience replay memory is a key component of D2SAC, as it allows the algorithm to handle the delay in receiving reward feedback. This is in contrast to traditional scheduling algorithms that require immediate utility feedback. Experience replay memory allows D2SAC to store experiences $(s_l,a_l,s_{l+1})$ and fill in missing reward $r_{l}$ at a later time before updating the AGOD network. Off-policy training is used to further improve the algorithm's ability to handle delayed feedback. Noted that, while the introduction of experience replay does bring a delay into the learning process, it does not impact the real-time performance in the decision process because the system's policy can be updated and used in real time, while learning takes place concurrently in an asynchronous manner.
	
	\textit{Double Critic Network.}
	During the policy improvement process, AGOD ${\pi}_{\boldsymbol{\theta}}$ is optimized by sampling mini-batches of transitions from experience replay memory $\mathcal{D}$. A double critic network is used as the Q function to reduce overestimation bias. Each critic network has its own set of parameters, denoted as $\boldsymbol{\phi}^1$ and $\boldsymbol{\phi}^2$, respectively. Both networks are updated independently using the same optimization target. During training, the Q-value estimate used to update the actor-network is the minimum of the two Q-value estimates from the two critic networks. This approach ensures that the actor-network is updated based on a conservative estimate of the Q-value function, promoting stable and efficient training.
	In contrast to the Q function $Q_{\boldsymbol{\phi}}(s_l,a_l)$ defined in the policy gradient theorem~\cite{sutton1999policy}, D2SAC~employs a more efficient Q function, denoted $Q_{\boldsymbol{\phi}}(s_l)$, where $\boldsymbol{\phi}=\{\boldsymbol{\phi}^1,\boldsymbol{\phi}^2\}$. Instead of only outputting the Q-value for a specific action, this Q function outputs a Q-value vector $\boldsymbol{q} \in \mathbb{R}^{|\mathcal{A}|}$ containing the Q-values of all possible actions $a_l\in\mathcal{A}$, i.e., $\boldsymbol{q}=Q_{\boldsymbol{\phi }}\left( s_l \right)=\min \left\{ Q_{\boldsymbol{\phi }^1}\left( s_l \right) ,Q_{\boldsymbol{\phi }^2}\left( s_l \right) \right\}$.
	
	\textit{Policy Improvement.}
	The Q-values $\boldsymbol{q}$ estimate the expected cumulative reward for each action at the current state $s_l$. Then, the actor can learn to maximize the expectation of $\boldsymbol{q}$ over all actions to improve the policy, which is expressed as:
	\begin{equation}
		\label{eq:max-q-expectation}
		\underset{\boldsymbol{\theta }}{\max}\,\, \pi _{\boldsymbol{\theta }}\left( s_l \right) ^TQ_{\boldsymbol{\phi }}\left( s_l \right).
	\end{equation}
	Maximizing \eqref{eq:max-q-expectation} encourages the current policy $\pi_{\boldsymbol{\theta}}$ to update in the direction where the actions with higher Q-values can increase their probabilities of being selected, while the others are suppressed. This maximization problem is solved using the gradient ascent algorithm, which can be transformed into a minimization problem expressed as
	\begin{equation}
		\label{eq:min-q-expectation}
		\underset{\boldsymbol{\theta }}{\min}\,\, -\pi _{\boldsymbol{\theta }}\left( s_l \right) ^TQ_{\boldsymbol{\phi }}\left( s_l \right).
	\end{equation}
	The standard gradient descent algorithm, such as Adam, can be used to solve this problem.
	Specifically, the gradient of \eqref{eq:min-q-expectation} with respect to the policy parameters $\boldsymbol{\theta }$ can be computed as the expectation over a mini-batch of transitions of size $b$ sampled from the experience replay memory $\mathcal{D}$ at the $e$-th training step, denoted by $\mathcal{B}_e$. Therefore, the gradient is given by
	\begin{equation}
		\mathbb{E}_{s_l\sim \mathcal{B}_e}\left[ -\nabla _{\boldsymbol{\theta }_e}\pi _{\boldsymbol{\theta }_e}\left( s_l \right) ^TQ_{\boldsymbol{\phi }_e}\left( s_l \right) \right],
	\end{equation}
	where $\boldsymbol{\theta }_e$ and $\boldsymbol{\phi }_e$ are the policy and Q-function parameters at the $e$-th training step, respectively. 
	The actor is then updated by performing gradient descent with respect to the above gradient, as follows,
	\begin{equation}
		\label{eq:update-actor}
		\boldsymbol{\theta }_{e+1}\gets \boldsymbol{\theta }_e-\eta_{\mathrm{a}}\cdot \left( \mathbb{E}_{s_l\sim \mathcal{B}_e}\left[ -\nabla _{\boldsymbol{\theta }_e}\pi _{\boldsymbol{\theta }_e}\left( s_l \right) ^TQ_{\boldsymbol{\phi }_e}\left( s_l \right) \right] \right),
	\end{equation}
	where $\eta_{\mathrm{a}}$ is the learning rate of the actor. By iteratively performing \eqref{eq:update-actor}, D2SAC~learns an optimal policy parameters that maximizes the sub-goal \eqref{eq:max-q-expectation}.
	
	\textit{Action Entropy Regularization.} To prevent the policy from becoming overly confident in certain actions and converging prematurely to a suboptimal solution, D2SAC~introduces an action entropy regularization term on the vanilla target \eqref{eq:max-q-expectation} to encourage exploration,
	\begin{gather}
		\label{eq:max-q-expectation-entropy}
		\underset{\boldsymbol{\theta }}{\max}\,\, \pi _{\boldsymbol{\theta }}\left( s_l \right) ^TQ_{\boldsymbol{\phi }}\left( s_l \right) +\alpha H\left( \pi _{\boldsymbol{\theta }}\left( s_l \right) \right)  
		\\
		\mathrm{s.t.}\,\, H\left( \pi _{\boldsymbol{\theta }}\left( s_l \right) \right) =-\pi _{\boldsymbol{\theta }}\left( s_l \right) ^T\log \pi _{\boldsymbol{\theta }}\left( s_l \right) 
	\end{gather}
	where $H\left( \pi _{\boldsymbol{\theta }}\left( s_l \right) \right)$ is the entropy of the action probability distribution $\pi _{\boldsymbol{\theta }}\left( s_l \right)$, and the temperature coefficient $\alpha$ controls the strength of the entropy term.
	Following the derivation process similar to \eqref{eq:max-q-expectation}-\eqref{eq:update-actor}, the update rule in \eqref{eq:update-actor} should add the gradient term of the entropy term $\nabla _{\boldsymbol{\theta }_e}H\left( \pi _{\boldsymbol{\theta }_e}\left( s_l \right) \right) $, as follow,
	\begin{equation}
		\label{eq:update-actor-soft}
		\boldsymbol{\theta }_{e+1}\gets \boldsymbol{\theta }_e-\eta _{\mathrm{a}}\cdot \mathbb{E}_{s_l\sim \mathcal{B}_e}\left[ \begin{array}{c}
			-\alpha \nabla _{\boldsymbol{\theta }_e}H\left( \pi _{\boldsymbol{\theta }_e}\left( s_l \right) \right)\\
			-\nabla _{\boldsymbol{\theta }_e}\pi _{\boldsymbol{\theta }_e}\left( s_l \right) ^TQ_{\boldsymbol{\phi }_e}\left( s_l \right)\\
		\end{array} \right].
	\end{equation}
	
	\textit{Q-function Improvement.}
	Ensuring accurate estimates of the Q-function $Q_{\boldsymbol{\phi }_e}\left( s_l \right) $ is crucial to the success of finding the optimal policy $\pi^*_{\boldsymbol{\theta}}$. Thus, $Q_{\boldsymbol{\phi }_e}\left( s_l \right) $ must be trained effectively. To update the Q-function, we minimize the Temporal Difference (TD) error between the Q-target $\hat{y}_e$ and the Q-eval $y_{e}^{i}$,
	\begin{align}
		\label{eq:td-error-loss}
		&\!\!\!\min_{\boldsymbol{\phi }^1,\boldsymbol{\phi }^2} && \mathbb{E}_{\left( s_l,a_l,s_{l+1},r_l \right) \sim \mathcal{B}_e}[ \sum_{i=1,2}{\left( \hat{y}_e-y_{e}^{i} \right) ^2} ],
		\\
		&\mathrm{s.t.} &&y_{e}^{i}=Q_{\boldsymbol{\phi }_{e}^{i}}\left( s_l,a_l \right),
		\\
		&&&\hat{y}_e=r_l+\gamma \left( 1-d_{l+1} \right) \hat{\pi} _{\boldsymbol{\hat{\theta}}_e}\left( s_{l+1} \right) ^T\hat{Q}_{\boldsymbol{\hat{\phi}}_e}\left( s_{l+1} \right).\label{eq:q-target}
	\end{align}
	Here, $Q_{\boldsymbol{\phi }_{e}^{i}}\left( s_l,a_l \right)$ denotes the Q-value corresponding to action $a_l$ output by $Q_{\boldsymbol{\phi }_{e}^{i}}\left( s_l \right)$, $\gamma$ represents the discount factor for future rewards, and $d_{l+1}$ is a 0-1 variable that represents the terminated flag. By updating the Q-function with the loss in \eqref{eq:td-error-loss}, we can improve the estimation accuracy of the Q-value.
	
	\textit{Target Networks.}
	In \eqref{eq:q-target}, ${\boldsymbol{\hat{\theta}}_e}$ and ${\boldsymbol{\hat{\phi}}_e}$ represent the parameters of the target actor $\hat{\pi}$ and the target critic $\hat{Q}$, respectively. The target networks $(\hat{\pi},\hat{Q})$ have the same network structure as the online networks $(\pi,Q)$, but their parameters $({\boldsymbol{\hat{\theta}}_e},{\boldsymbol{\hat{\phi}}_e})$ are frozen during gradient descent and are updated slowly through a soft update mechanism, which is defined as
	\begin{equation}
		\label{eq:soft-update}
		\begin{aligned}
			\boldsymbol{\hat{\theta}}_{e+1}&\gets \tau \boldsymbol{\theta }_e+(1-\tau )\boldsymbol{\hat{\theta}}_e,
			\\
			\boldsymbol{\hat{\phi}}_{e+1}&\gets \tau \boldsymbol{\phi }_e+(1-\tau )\boldsymbol{\hat{\phi}}_e,
		\end{aligned}    
	\end{equation}
	The hyperparameter $\tau \in (0,1]$ determines the update rate of the target network. A smaller value of $\tau$ leads to slower updates, while a larger value results in more rapid updates. By controlling $\tau$, the stability of the target network can be maintained. Finally, the D2SAC algorithm iteratively performs $E$ steps of policy and Q-function improvement until convergence is achieved. This results in near-optimal policy parameters $\boldsymbol{\theta}^*$ that maximize the cumulative reward in \eqref{eq:cumulative-reward}, which, in turn, maximizes the ultimate utility target in \eqref{eq:utility-target}.
	
	\subsection{Optimization Goal}
	Like most DRL tasks in communication and networking, the scheduling task is both online and discrete, making labeled actions unavailable for calculating the MSE loss. Moreover, the goal of D2SAC~is to maximize the Q-value, not to reconstruct an action probability distribution that does not exist. While the authors in \cite{wang2022diffusion} introduced a similar loss, called behavior cloning loss, for offline DRL tasks using imitation learning, it is impractical to obtain open datasets for online communication scheduling tasks. Additionally, approaches designed for general continuous control tasks \cite{wang2022diffusion,janner2022planning} cannot be applied in environments with discrete action spaces. In summary, the optimization goal of D2SAC~only needs to consider the policy loss and the action entropy loss, as defined in \eqref{eq:max-q-expectation-entropy}. Thus, we present the overview of our D2SAC~algorithm is then presented in {\textbf{Algorithm~\ref{alg:d2sac}}}. In the experiment part, we show that doing this way achieves excellent performance in various online and discrete-action tasks.

	\begin{algorithm}[t]
		\caption{{\small D2SAC: Deep Diffusion Soft Actor Critic}}\label{alg:d2sac} 
		{\small \begin{algorithmic}[1] 
				\State{Initialize policy parameters $\boldsymbol{\theta}$, Q-function parameters $\boldsymbol{\phi}$, target network parameters $\boldsymbol{\hat{\theta}}\gets \boldsymbol{\theta }$, $\boldsymbol{\hat{\phi}}\gets \boldsymbol{\phi }$, and replay buffer $\mathcal{D}$;}
				\For{the training step $e=1$ to $E$}
				\For{the number of collected transitions $c=1$ to $C$}
				\State{Observe state $s$ and initialize a random normal distribution $\boldsymbol{x}_T\sim\mathcal{N}(\boldsymbol{0},\mathbf{I})$;}
				\For{the {\color{black} denoising step} $t=T$ to 1}
				\State{Infer and scale a denoising distribution $\tanh \left( \boldsymbol{\epsilon }_{\boldsymbol{\theta }}(\boldsymbol{x}_t,t,s_l) \right)$ using a deep neural network;}
				\State{Calculate the mean $\boldsymbol{\mu }_{\boldsymbol{\theta }}$ of the reverse transition distribution $p_{\boldsymbol{\theta }}\left( \boldsymbol{x}_{t-1}|\boldsymbol{x}_t \right)$, as defined in \eqref{eq:p-transition} and \eqref{eq:reverse-mean};}
				\State{Calculate the distribution $\boldsymbol{x}_{t-1}$ using the reparameterization trick by \eqref{eq:xt-update-rule};}
				\EndFor
				\State{Calculate the probability distribution of $\boldsymbol{x}_0$ using \eqref{eq:softmax-output} and select action $a$ at random based on it.}
				\State{Execute action $a$ in the environment, and observe the next state $s^\prime$ and reward $r$;}
				\State{Store the transition $(s, a, s^\prime, r)$ in the replay buffer $\mathcal{D}$;}
				\EndFor
				\State{Sample a batch of transitions $\mathcal{B}=\{(s, a, s^\prime, r)\}$ from the replay buffer $\mathcal{D}$;}
				\State{Update the policy parameters $\boldsymbol{\theta}$ using $\mathcal{B}$ by \eqref{eq:update-actor-soft};}
				\State{Update the Q-function parameters $\boldsymbol{\phi}$ using $\mathcal{B}$ by one step of gradient descent to minimize \eqref{eq:td-error-loss};}
				\State{Update the target networks $\boldsymbol{\hat{\theta}},\boldsymbol{\hat{\phi}}$ using \eqref{eq:soft-update};}
				\EndFor
				\State{\Return{a AGOD-based policy $\pi^*$ with well-trained parameters $\boldsymbol{\theta}^*$;}}
		\end{algorithmic}}
	\end{algorithm}
	
	\subsection{Complexity Analysis}
	The computational complexity of D2SAC~is $\mathcal{O}\left( E\left[ CV+TC|\boldsymbol{\theta }|+\left( b+1 \right) \left( |\boldsymbol{\theta }|+|\boldsymbol{\phi }| \right) \right] \right)$. This complexity can be divided into two parts:
	\begin{itemize}
		\item Trajectory Collection: $\mathcal{O}(EC\left( V+T|\boldsymbol{\theta }| \right) )$. Throughout the $E$ training steps, $C$ trajectories are collected at each training step, resulting in a cumulative overhead of $\mathcal{O}(ECV)$ for the environment interaction. Furthermore, for each trajectory sampling, an additional overhead of $\mathcal{O}(T|\boldsymbol{\theta}|)$ is incurred due to the reverse diffusion process, which involves $T$ {\color{black} denoising step} of neural network inference.
		\item Parameter Updates: $\mathcal{O}(E\left( b+1 \right) \left( |\boldsymbol{\theta }|+|\boldsymbol{\phi }| \right) )$. This term is composed of three parts, i.e., $\mathcal{O}(bE|\boldsymbol{\theta}|)$ for policy improvement, $\mathcal{O}(bE|\boldsymbol{\phi}|)$ for Q-function improvement, and $\mathcal{O}\left( E\left( |\boldsymbol{\theta }|+|\boldsymbol{\phi }| \right) \right)$ for target network updates. Here, $b$ represents the batch size, and $|\boldsymbol{\theta }|$ and $|\boldsymbol{\phi }|$ are the number of parameters in the policy and Q-function networks, respectively.
	\end{itemize}
	The space complexity of D2SAC~is $\mathcal{O}\left( 2\left( |\boldsymbol{\theta }|+|\boldsymbol{\phi }| \right) +D\left( 2|\mathcal{S}|+|\mathcal{A}|+1 \right) \right)$. This includes storage for the policy and Q-function networks, as well as their target networks, which is $\mathcal{O}\left( 2\left( |\boldsymbol{\theta }|+|\boldsymbol{\phi }| \right) \right)$. Additionally, we need to store the trajectory experiences, which consist of $D$ transitions, each containing two state tuples of dimension $|\mathcal{S}|$, an action tuple of dimension $|\mathcal{A}|$, and a reward scalar. In summary, D2SAC~has the same space complexity as SAC, but its computational complexity increases by $\mathcal{O}\left( EC|\boldsymbol{\theta }|\left( T-1 \right) \right)$ due to the additional $T$ {\color{black} denoising step} in the reverse diffusion process. However, the increase in computational complexity helps to achieve higher performance and faster convergence, as demonstrated in Table \ref{tab:rl-comparison-with-benchmarks}.
	
	\section{Experiments and Insights}\label{Sexp}
	In this section, we comprehensively evaluate the AGOD-based D2SAC algorithm and demonstrate its superior performance compared with existing methods. Our analyses also provide valuable insights into the use of diffusion-based DRL in discrete action spaces.
	
	\subsection{Experimental Setup}
	\textbf{Experimental Platform.}
	Our experiments were conducted with an NVIDIA GeForce RTX 3090 GPU with 24 GB of memory and an AMD Ryzen 9 5950X 16-Core processor with 128 GB of RAM. The workstation was running Ubuntu 16.04 LTS operating system and utilized PyTorch 1.13 along with the CUDA toolkit 11.6 and cuDNN 8.0. We packaged our software environment and dependencies into a Docker image to ensure reproducibility.
	
	\textbf{Environment Details.}
	We train an agent to assign Metaverse users' AIGC tasks to wireless ASPs in a simulation environment with $20$ ASPs. Each ASP had a random resource capacity $\mathcal{T}$, which represented the total available {\color{black} denoising step} for the diffusion process and ranged from $400$ to $1000$. 
	{\color{black}We use RePaint\footnote{Github: \url{https://github.com/andreas128/RePaint}}~\cite{lugmayr2022repaint} as the AIGC model and PyTorch Image Quality (PIQ)\footnote{Github: \url{https://github.com/photosynthesis-team/piq}}~\cite{kastryulin2022piq} as the human-aware content quality assessment function. Note that the quality of AIGC services that different ASPs provide can vary, as depicted in Fig.~\ref{aaas} (Part C). 
		A linear function parameterized by $A_x$, $A_y$, $B_x$, and $B_y$ was determined based on tests using the real image dataset CelebA-HQ~\cite{liu2015deep} to model the quality of images generated by an AIGC model~\cite{du2023enabling}. 
		To simulate the varying capabilities of different ASPs, we set $A_x\in[0,100], A_y\in[0, 0.5], B_x\in[150,250]$, and $B_y\in[0.5,1]$.
		Our simulations involved $1000$ Metaverse users submitting multiple AIGC task requests to the ASPs at different times. Given the unpredictable nature of user behavior, each request was assumed to require a random amount of resources $T$ (i.e., the number of denoising steps) ranging from $100$ to $250$. 
		The arrival of user tasks was modeled as a Poisson process\footnote{The use of a Poisson process in modeling the arrival of user tasks in our experiments is motivated by its wide acceptance and prevalence~\cite{ko2018wireless,chung2005performance}. Fortunately, leveraging the inherent adaptability of reinforcement learning, our proposed method offers flexibility, capably managing task arrivals that follow various distributions.}, i.e.,
		$P(k; \lambda) = \frac{\lambda^k e^{-\lambda}}{k!}$, where $\lambda=0.001$ is the average arrival rate, and $J=1000$ is the number of tasks that arrive in the time interval of $1\times10^6$ seconds}. To manage the ASPs and user task requests, we implemented a swarm manager that allocated task requests to ASPs based on the action decided by D2SAC. We monitored the operation status to measure the performance.
	
	\textbf{Model Design.}
	D2SAC~employs the diffusion model-based AGOD as the core of the actor network and uses two critic networks with the same structure to mitigate the problem of overestimation. Table \ref{tab:actor-critic-structure} shows the detailed configurations of the actor and critic networks.
	\begin{table}[htbp]
		\centering
		\begin{minipage}[t]{0.35\textwidth}
			\centering
			\caption{The structure of actor and critic networks}
			\label{tab:actor-critic-structure}
			\footnotesize
			\renewcommand\arraystretch{1.25}
			\begin{tabular}{c|c|c|c}
				\hline
				\multicolumn{1}{c|}{\textbf{Networks}} & \textbf{Layer} & \textbf{Activation} & \textbf{Units} \\
				\hline
				\multirow{7}{*}{Actor} & SinusoidalPosEmb & - & 16 \\
				& FullyConnect & Mish & 32 \\
				& FullyConnect & - & 16 \\
				& Concatenation & - & - \\
				& FullyConnect & Mish & 256 \\
				& FullyConnect & Mish & 256 \\
				& FullyConnect & Tanh & 20 \\
				\hline
				\multirow{3}{*}{Critic} & FullyConnect & Mish & 256 \\
				& FullyConnect & Mish & 256 \\
				& FullyConnect & - & 20 \\
				\hline
			\end{tabular}
		\end{minipage}%
		\hfill
		\begin{minipage}[t]{0.5\textwidth}
			\centering
			\caption{Summary of Training Hyperparameters}
			\label{tab:hyper-params-setup}
			\footnotesize
			\renewcommand\arraystretch{1.2}
			\begin{tabular}{c l c}
				\hline
				\textbf{Symbol} &
				\textbf{Description} &
				\textbf{Value} \\
				\hline
				$\eta_{\mathrm{a}}$ & Learning rate of the actor network & $1\times10^{-4}$ \\
				$\eta_{\mathrm{c}}$ & Learning rate of the critic networks & $1\times10^{-3}$ \\
				$\alpha$ & Temperature of action entropy regularization & 0.05 \\
				$\tau$ & Weight of soft update & 0.005 \\
				$b$ & Batch size & 512 \\
				$\lambda$ & Weight decay & $1\times10^{-4}$ \\
				$\gamma$ & Discount factor to accumulate rewards & 0.95 \\
				$T$ & {\color{black} Denoising steps} for the diffusion model & 5 \\
				$D$ & Maximum capacity of the replay buffer & $1\times10^6$ \\
				$E$ & Total number of training steps & 1000 \\
				$C$ & Number of transitions per training step & 1000 \\
				\hline
			\end{tabular}
		\end{minipage}
	\end{table}
	The actor-network in D2SAC~not only predicts the denoised distribution from a random normal distribution and the current state but also includes {\color{black} denoising step} encodings, i.e., Sinusoidal position embeddings~\cite{vaswani2017attention}, to capture temporal information about the diffusion process. This helps the actor-network better understand the relationships between each step in the diffusion chain.
	The backbone of the actor-network consists of three fully-connected layers that use the Mish activation function, except for the final layer, which employs the Tanh activation to normalize its outputs. The critic networks share a similar structure with the actor-network, consisting of three fully-connected layers with Mish activations. However, the final layer of the critic networks produces Q values for actions without any activation function.
	The actor and critic networks are trained by using the Adam optimizer~\cite{kingma2014adam}, with a learning rate of $\eta_\mathrm{a}=0.0001$ for the actor-network and $\eta_\mathrm{c}=0.001$ for the critic networks. A weight decay rate of $\lambda=0.0001$ was employed to regularize model weights and promote learning more policies. The target networks mirrored the structures of the online networks to reduce variance during the learning process. By default, we set the update rate $\tau=0.005$ for soft updating the target networks, as defined in \eqref{eq:soft-update}.
	The detailed settings for other training hyperparameters in our experiments are summarized in Table \ref{tab:hyper-params-setup}.
	
	\textbf{Benchmarks.} In our experiments, we compare the D2SAC~with seven well-known DRL benchmarks, including DQN~\cite{mnih2015human}, DRQN~\cite{hausknecht2015deep}, Prioritized-DQN~\cite{schaul2015prioritized}, Rainbow~\cite{hessel2018rainbow}, REINFORCE~\cite{williams1992simple}, PPO~\cite{schulman2017proximal}, and SAC~\cite{haarnoja2018soft}. Specifically, DQN, DRQN, Prioritized-DQN, and Rainbow are value-based methods suited for optimization problems with discrete action spaces. The other algorithms are policy-based and were evaluated in the discrete action space to ensure fair comparisons. Despite similarities to SAC, D2SAC replaces the actor-network with diffusion model-based AGOD. In the following experiments, we demonstrate the superiority of D2SAC over these benchmarks and present interesting insights. In addition to these advanced DRL benchmarks, we implement four heuristic policies:
	\begin{itemize}
		\item {\textit{\textbf{Random Policy.}}} The random policy assigns incoming tasks to ASPs randomly without considering available resources, task processing time, or other constraints. This policy serves as the lower-bound baseline for scheduling performance.
		\item {\textit{\textbf{Round Robin Policy.}}} The Round Robin policy assigns tasks to ASPs in cyclical order. This policy can produce favorable schedules when tasks are well-balanced. However, it may not perform optimally without significant differences among tasks~\cite{garcia2021enhancing}.
		\item {\textit{\textbf{Crash Avoid Policy.}}} The Crash Avoid policy assigns tasks to ASPs based on their available resources. ASPs with more resources are given priority in task assignments to prevent overloading.
		\item {\textit{\textbf{Prophet Policy.}}} We assume that the scheduler knows in advance every utility that the ASP will bring to every user before assigning tasks. In this case, the prophet policy provides an upper bound on the performance of human-centric services, by assigning tasks to ASPs with the highest utility while avoiding crashes. However, this policy uses the unknown utility function as prior information before tasks are assigned, which is not feasible in the real world.
	\end{itemize}
	
	\subsection{Numerical Results}
	\textbf{Leading Performance.}
	For the ASP selection problem, we summarize the best performance achieved by the proposed D2SAC and 11 benchmark policies in Table \ref{tab:rl-comparison-with-benchmarks}, in terms of cumulative reward, training time, and convergence speed. Each experiment was run for $E=1000$ training steps and in a total of $1\times 10^6$ environment steps. To assess the time efficiency and convergence speed, we used the Crash Avoid policy as the baseline. We recorded the time and steps taken by each policy to reach the baseline reward. The {\textit{time to baseline}} and {\textit{step to baseline}} refer to the time and the number of training steps when the test reward reaches that of the Crash Avoid policy, respectively.
	
\begin{figure*}[t]
		\centering
		\begin{subfigure}[{\color{black}D2SAC~vs DQN, DRQN, Prioritized-DQN, and Rainbow}]{\includegraphics[width=0.3\linewidth]{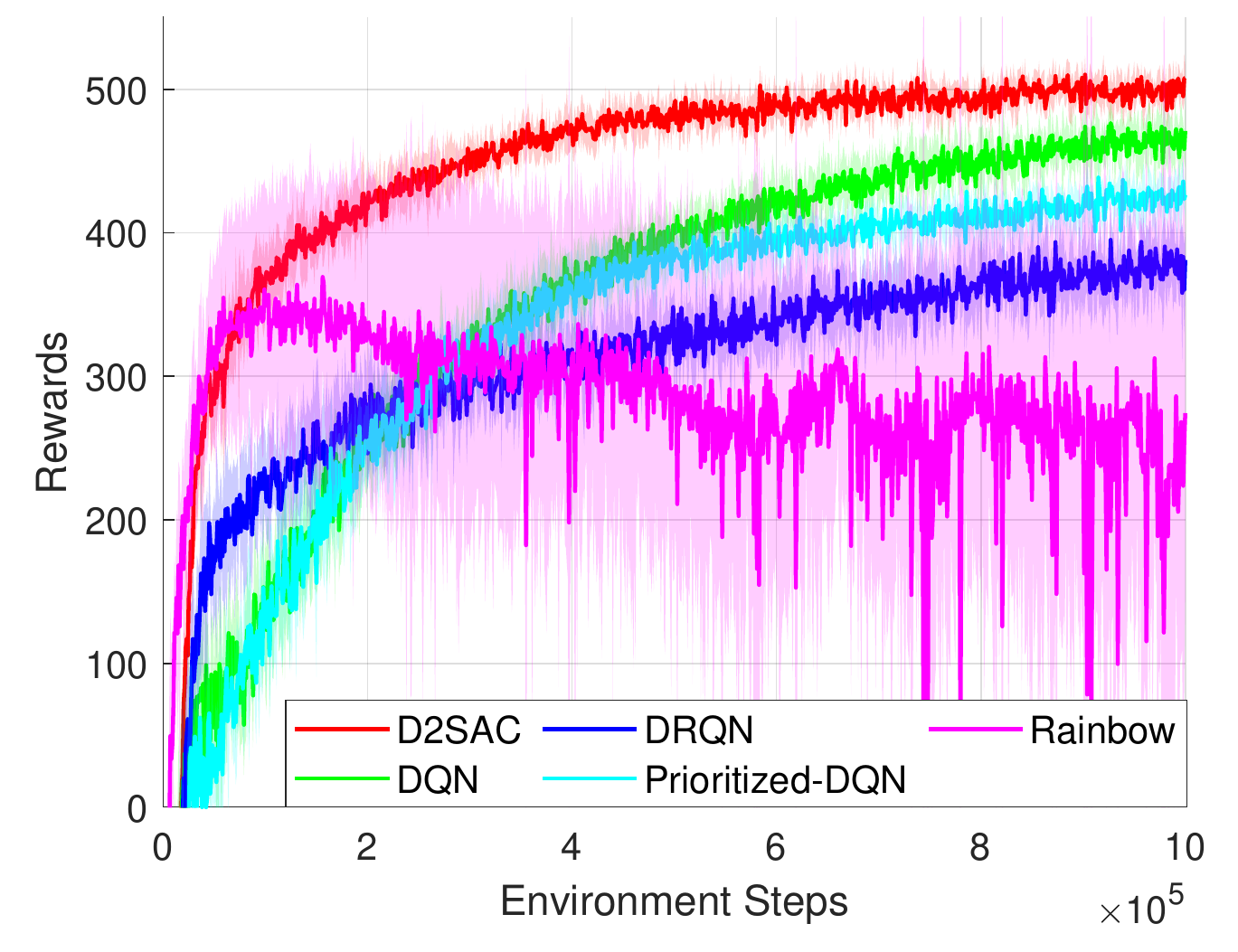}}
		\end{subfigure}
		\hfill
		\begin{subfigure}[{\color{black}D2SAC~vs REINFORCE, PPO, and SAC}]{\includegraphics[width=0.3\linewidth]{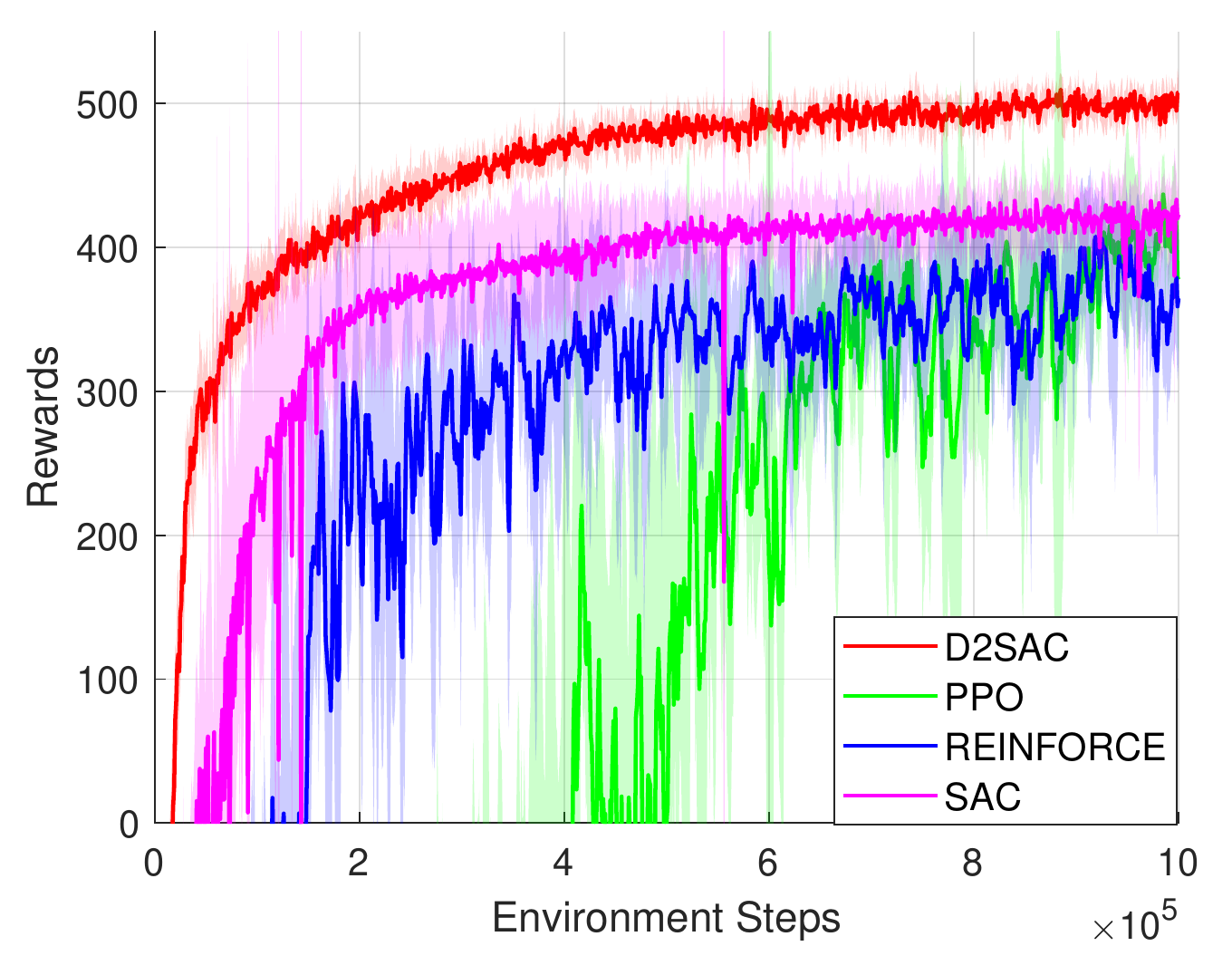}}
		\end{subfigure}
		\hfill
		\begin{subfigure}[{\color{black}D2SAC~vs Prophet, Round Robin, Crash Avoid, and Random policies}]{\includegraphics[width=0.29\linewidth]{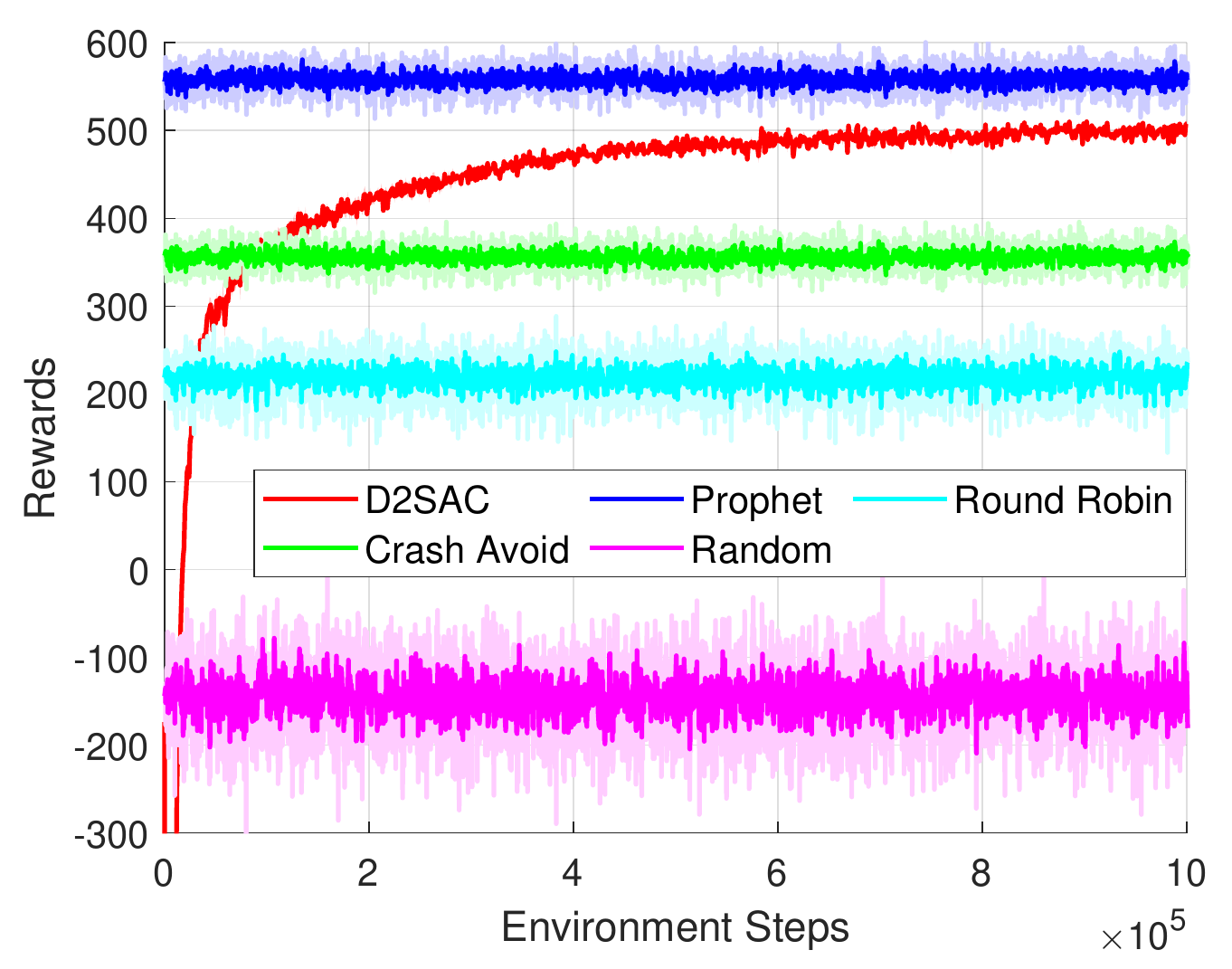}}
		\end{subfigure}
		\caption{{\color{black}Comparison of test reward curves of D2SAC~and benchmarks.}}
		\label{fig:comparison-of-curves}
	\end{figure*}

\begin{figure*}
		\centering
		\includegraphics[width=1\linewidth]{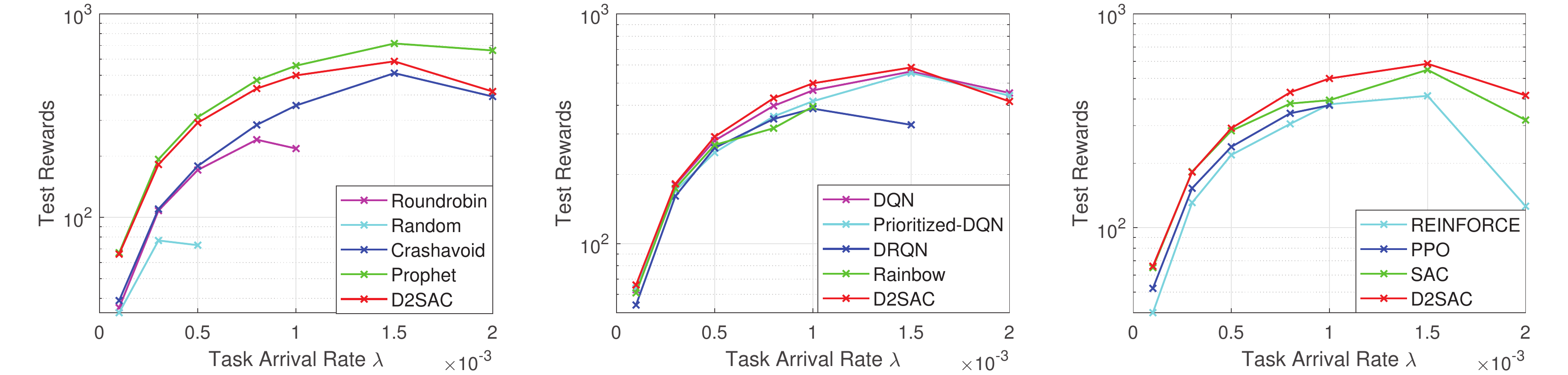}
		\caption{{\color{black}The cumulative test rewards over different task arrival rates. Negative reward values are not displayed.}}
		\label{fig:ratediff}
	\end{figure*}

	\begin{table*}
		\centering
		\caption{Performance Comparisons of D2SAC~and Benchmarks (Totally 1000 Steps).}
		\label{tab:rl-comparison-with-benchmarks}
		\footnotesize
		\renewcommand\arraystretch{1.25}
		\begin{tabular}{c|c|ccccc}
			\hline
			\multicolumn{2}{c|}{\textbf{Policy}} & \textbf{Train Reward} & \textbf{Test Reward} & \textbf{Toal Time / h} & \textbf{Time to Baseline / h} & \textbf{Step to Baseline} \\ 
			\hline\hline
			\multirow{4}{*}{Heuristic} & Random & -21 & -35  & 0.74 & $\infty$ & $\infty$ \\ 
			& Round Robin & 273 & 280 & 0.76 & $\infty$ & $\infty$ \\ 
			& Crash Avoid & 389 & 400 & 0.77 & 0.0 & 0 \\ 
			& Prophet & 604 & 596 & $\infty$ & $\infty$ & $\infty$ \\ 
			\hline\hline
			\multirow{7}{*}{DRL} & DQN & 418 & 503 & 1.9 & \textbf{0.9} & 470 \\
			& Prioritized-DQN & 386 & 460 & 1.8 & 1.0 & 470 \\ 
			& DRQN & 384 & 430 & 2.9 & 2.0 & 700 \\ 
			& REINFORCE & 395 & 463 & \textbf{1.1} & \textbf{0.9} & 850 \\ 
			& PPO & 353 & 481 & \textbf{1.1} & 1.1 & 950 \\ 
			& Rainbow & 414 & 450 & 2.6 & 2.2 & \{130,850\} \\ 
			& SAC & 418 & 436 & 2.9 & 1.2 & 430 \\ 
			\hline\hline
			\textbf{Ours} & \textbf{D2SAC} & \textbf{528} & \textbf{537} & 7.0 & 1.3 & \textbf{190} \\
			\hline
		\end{tabular}
	\end{table*}
	The DRL-based policies outperformed the Crash Avoid policy, as shown in Table \ref{tab:rl-comparison-with-benchmarks}. However, there is still a significant variation in performance among different policies. REINFORCE and PPO, have relatively short training times but produce subpar results, while DQN and our proposed D2SAC require longer training times but achieve better performance. Notably, D2SAC stands out in the comparison, delivering the highest training and test rewards, achieving the baseline reward after only 190 training steps, and a relatively fast training time of $1.3$ hours. The superior performance of D2SAC can be attributed to its use of the diffusion model-based AGOD, which enhances its capability to capture complex patterns and relationships in the observations.
	{\color{black} The performance of D2SAC in comparison with the other policies is further evaluated in Fig.~\ref{fig:comparison-of-curves}. Furthermore, Fig.~\ref{fig:ratediff} shows the test rewards of different policies under various task arrival rates, which verifies the robustness of the proposed D2SAC.}
	More importantly, we compare the D2SAC with other DRL algorithms through various standard control tasks in the Gym environment, as presented in Table \ref{tab:comparison-on-general}. These results demonstrate the superior characteristics of D2SAC in terms of \textit{high-performance}, \textit{time-efficient}, and \textit{fast-converging}, positioning it as the top choice for discrete action scenarios such as the ASP selection in wireless edge networks.
	
	\begin{table*}
		\centering
		\caption{Accumulated Reward Comparisons on General Benchmark Tasks.}
		\label{tab:comparison-on-general}
		\footnotesize
		\renewcommand\arraystretch{1.25}
		\begin{tabular}{c|c|cccc}
			\hline
			\multicolumn{2}{c|}{\textbf{Policy}} & \textbf{Acrobot-v1} & \textbf{CartPole-v1} & \textbf{CoinRun-v0} & \textbf{Maze-v0} \\
			\hline\hline
			\multirow{7}{*}{DRL} & DQN & -81.81 $\pm$ 17.19 & 499.80 $\pm$ 0.14 & 6.00 $\pm$ 4.90 & 3.00 $\pm$ 4.58 \\
			& Prioritized-DQN & -105.20 $\pm$ 14.74 & 498.70 $\pm$ 1.43 & 5.00 $\pm$ 5.00 & 2.00 $\pm$ 4.00 \\ 
			& DRQN & -82.26 $\pm$ 14.34 & 132.50 $\pm$ 69.79 & $-$ & $-$ \\ 
			& REINFORCE & -104.80 $\pm$ 14.51 & 500.00 $\pm$ 0.00 & 0.00 $\pm$ 0.00 & 0.00 $\pm$ 0.00 \\ 
			& PPO & -77.22 $\pm$ 8.45 & 499.90 $\pm$ 0.33 & 0.00 $\pm$ 0.00 & 2.00 $\pm$ 4.00 \\ 
			& Rainbow & -158.10 $\pm$ 55.48 & 478.30 $\pm$ 29.28 & 5.00 $\pm$ 5.00 & 2.00 $\pm$ 4.00 \\ 
			& SAC & -121.00 $\pm$ 35.31 & 500.00 $\pm$ 0.00 & 10.00 $\pm$ 0.00 & 3.00 $\pm$ 4.58 \\ 
			\hline\hline
			\multirow{8}{*}{Online~\cite{pmlr-v119-cobbe20a,rl-zoo}} & A2C & -86.62 $\pm$ 25.10 & 499.90 $\pm$ 1.67 & $-$ & $-$ \\
			& ACER & -90.85 $\pm$ 32.80 & 498.62 $\pm$ 23.86 & $-$ & $-$ \\
			& ACKTR & -91.28 $\pm$ 32.52 & 487.57 $\pm$ 63.87 & $-$ & $-$ \\
			& PPO2 & -85.14 $\pm$ 26.27 & 500.00 $\pm$ 0.00 & $-$ & $-$ \\
			& DQN & -88.10 $\pm$ 33.04 & 500.00 $\pm$ 0.00 & $-$ & $-$ \\
			& TRPO & $-$ & 485.39 $\pm$ 70.51 & $-$ & $-$ \\
			& PPO + IMPALA & $-$ & $-$ & 8.95 & \textbf{9.88} \\
			& Rainbow + IMPALA & $-$ & $-$ & 5.50 & 4.24 \\
			\hline\hline
			\textbf{Ours} & \textbf{D2SAC} & \textbf{-70.77} $\pm$ \textbf{4.12} & \textbf{500.00} $\pm$ \textbf{0.00} & \textbf{10.00} $\pm$ \textbf{0.00} & 7.00 $\pm$ 4.58 \\
			\hline
		\end{tabular}
	\end{table*}
	
	\textbf{Understand the Learning Process.}
	To gain insights into the learning process of D2SAC, we compared it against conventional heuristic policies in subfigure (c) of Fig.~\ref{fig:comparison-of-curves}. These heuristic policies rely on simple or random rules to make action decisions. While these policies are easy to interpret, they are suboptimal. D2SAC and other DRL-based policies can adapt to changing environments and maximize performance over time. D2SAC interacts with the environment during the learning process by taking action and learning from feedback rewards. This information is then used to improve its decision-making process, i.e., the AGOD network, leading to continuous performance enhancement.
	\begin{figure}[t]
		\centering
		\includegraphics[width=0.9\linewidth]{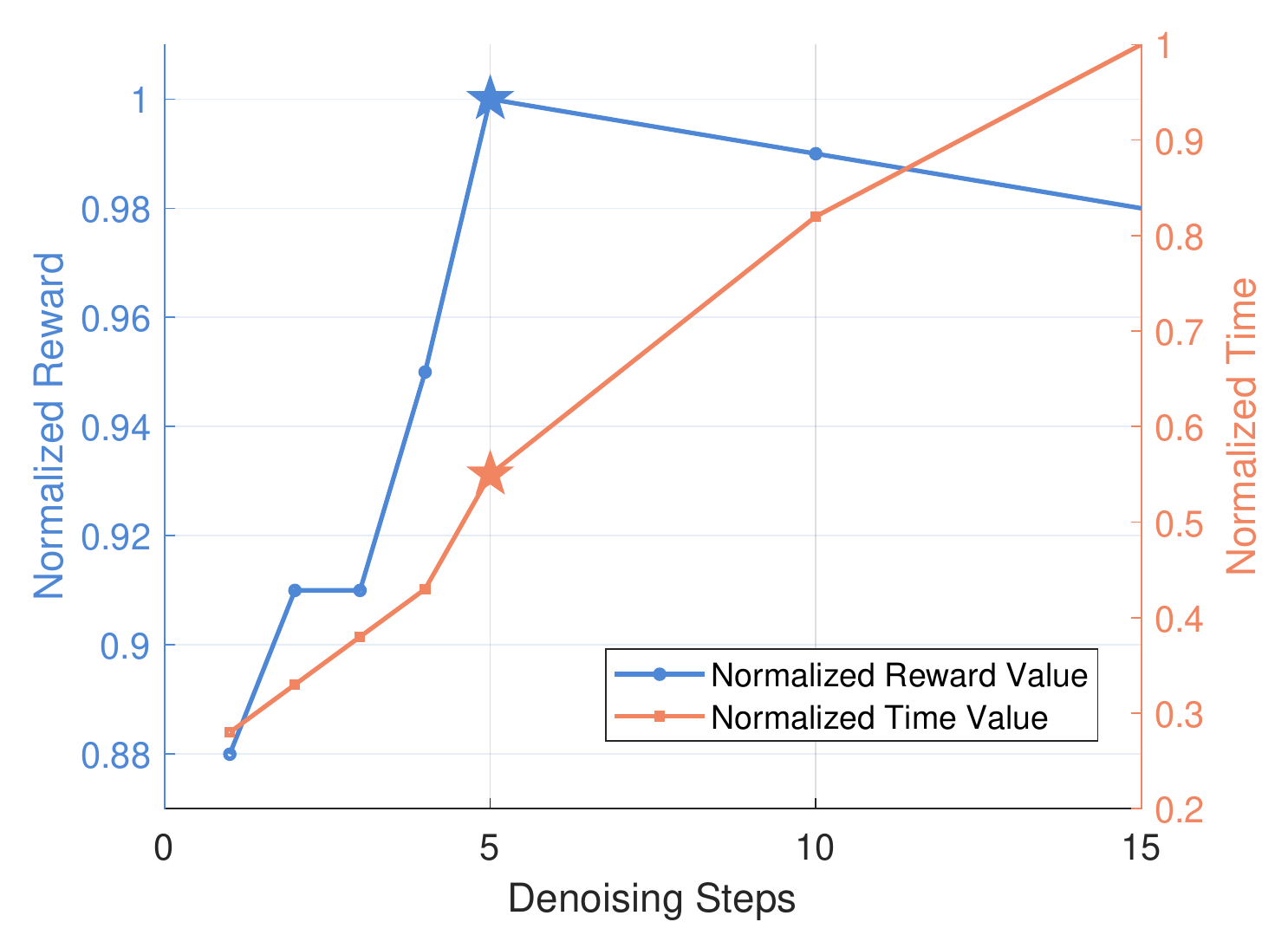}
		\caption{Denoising step impact on reward and training time, normalized to their maximum value.}
		\label{fig:denoisingstep}
	\end{figure}
	{\color{black}D2SAC begins with a random policy, progressively learning the optimal one through trial and error in the environment. It outperformed the Round Robin policy after about $45$ training steps and exceeded the Crash Avoid baseline by $80$ steps, showing superior load-balancing and crash-prevention abilities. Initially, D2SAC prioritized task completion over utility optimization. Over time, its policy refined and approached the theoretical upper limit, akin to the prophet policy. This progression demonstrates that D2SAC can maintain its Crash Avoiding capability while maximizing user utility.}
	
	\begin{table*}
		\centering
		\caption{Task Performance Comparisons of D2SAC~and Benchmarks}
		\label{tab:task-comparison-with-benchmarks}
		\footnotesize
		\renewcommand\arraystretch{1.25}
		\begin{tabular}{c|c|cccc}
			\hline
			\multicolumn{2}{c|}{\textbf{Policy}} & \textbf{Finished Rate}\tablefootnote{The sum of finish rate and crash rate is not strictly 1, because some tasks are still running.} & \textbf{Obtained Utility} & \textbf{Crashed Rate} & \textbf{Lost Utility} \\ 
			\hline\hline
			\multirow{4}{*}{Heuristic} & Random & 70.2\% & 215 & 27.7\% & 93 \\ 
			& Round Robin & 90.3\% & 309 & 7.6\% & 32 \\ 
			& Crash Avoid & 97.7\% & 357 & 0\% & 0 \\ 
			& Prophet & 97.7\% & 548 & 0\% & 0 \\ 
			\hline\hline
			\multirow{7}{*}{DRL} & DQN & \textbf{97.7\%} & 479 & \textbf{0.0\%} & \textbf{0} \\
			& Prioritized-DQN & \textbf{97.7\%} & 433 & \textbf{0.0\%} & \textbf{0} \\ 
			& DRQN & 94.3\% & 433 & 3.8\% & 17 \\ 
			& REINFORCE & 95.8\% & 458 & 1.9\% & 10 \\ 
			& PPO & 97.0\% & 457 & 0.7\% & 4 \\ 
			& Rainbow & \textbf{97.7\%} & 419 & \textbf{0.0\%} & \textbf{0} \\ 
			& SAC & 94.3\% & 412 & 3.5\% & 11 \\ 
			\hline\hline
			\textbf{Ours} & \textbf{D2SAC} & 96.6\% & \textbf{494} & 1.1\% & 5 \\
			\hline
		\end{tabular}
	\end{table*}
	
	\textbf{New and Advanced Abilities.}
	The results presented in Table \ref{tab:task-comparison-with-benchmarks} offer a comprehensive comparison of several metrics, including finish rate, obtained utility, crash rate, and lost utility. The finished and crash rates indicate the percentage of completed and crashed tasks, respectively. The obtained utility is the total rewards, while the lost utility reflects the rewards lost due to task crashes. The data in Table \ref{tab:task-comparison-with-benchmarks} indicate that all DRL-based policies outperform the heuristic policies regarding obtained utility and provide competitive benchmarks to our D2SAC. This observation is consistent with the findings from Table \ref{tab:rl-comparison-with-benchmarks}. However, policies such as REINFORCE, PPO, and the proposed D2SAC, which achieved high utility, still experienced a near-zero crash rate. This highlights the trade-offs required to maximize utility, as some crashes are inevitable. Conversely, policies such as Rainbow, which focused on zero crashes, suffered from the lower utility.
	Among the DRL-based policies, DQN achieved the highest utility with no crashes. However, D2SAC outperformed DQN regarding utility, indicating that D2SAC learned to prioritize tasks by estimating their values and selectively discarding low-value tasks to reserve resources for high-value tasks. This insight is further evident in the comparison between PPO and D2SAC, where D2SAC crashed $1.1\%$ of tasks with a lost utility of 5, while PPO crashed $0.7\%$ of tasks with a lost utility of 4. This feature is precious in real-world scheduling systems where resource allocation is critical. However, when avoiding crashes is of utmost importance, DQN might be a better option.

	\textbf{No Longer Large {\color{black}Denoising Step}.} The diffusion chain in diffusion-based generation models refers to the sequential spread of information from one state to another, with the length of the chain represented by the {\color{black} denoising step} $T$. Selecting an appropriate value for $T$ involves a trade-off between computational efficiency and accuracy. To ensure accuracy, a large value of $T$ is recommended, but this comes at the cost of longer computation times. However, a small value of $T$ reduces computation time but can increase the risk of instability and numerical errors. In a recent study \cite{ho2020denoising}, a value of $T=500$ was found to strike a balance between accuracy and computational efficiency.
	
	However, in D2SAC, the relationship between the {\color{black} denoising step}, reward, and computational time did not follow the above rule. Specifically, in Fig.~\ref{fig:denoisingstep}, we vary the {\color{black} denoising step} $T\in{1,2,3,4,5,10,15}$. We observe that the reward first increased, but then decreased as the number of {\color{black} denoising step} increased, while the training time consistently increased. This finding suggests that there is an optimal {\color{black} denoising step} at the inflection point of the reward curve, which appears to be $T=5$. Moreover, we discovered that the optimal {\color{black} denoising step} was significantly fewer than the one used in \cite{ho2020denoising}, indicating that the trade-off between learning performance and computational efficiency was no longer present. Thus, taking a small $T$ can achieve a satisfying reward while maintaining high computational efficiency.
	
\begin{figure}[htbp]
		\centering
		\begin{minipage}[t]{0.5\textwidth}
			\centering
			\includegraphics[width=1\linewidth]{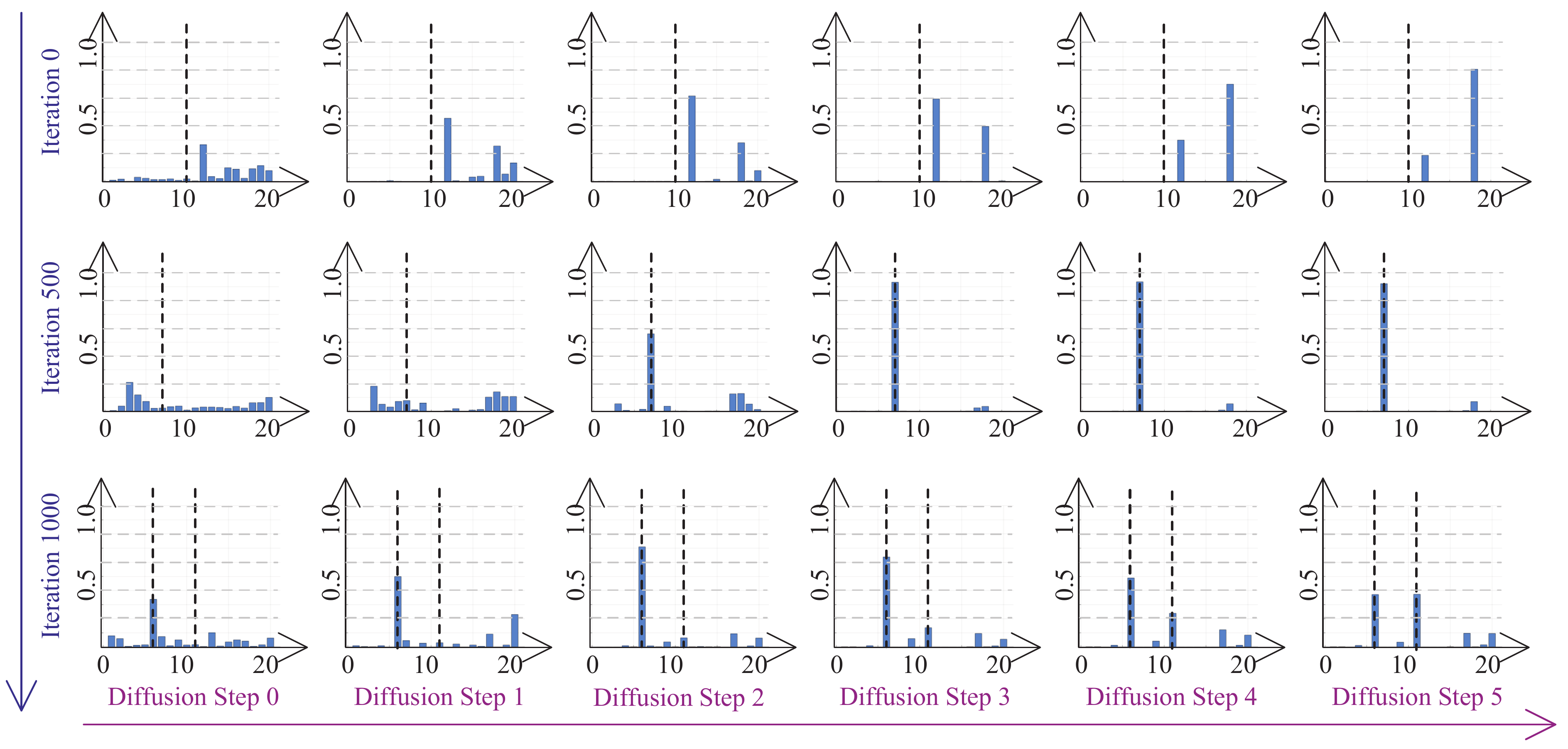}
			\vspace{-0.5cm}
			\caption{Illustration of the ``moving" action probability distribution during the reverse diffusion process, i.e., AGOD algorithm. The vertical dotted lines represent the optimal action(s) predicted by the prophet policy.}
			\label{fig:diffusion}
		\end{minipage}%
		\hfill
		\begin{minipage}[t]{0.45\textwidth}
			\centering
			\includegraphics[width=0.9\linewidth]{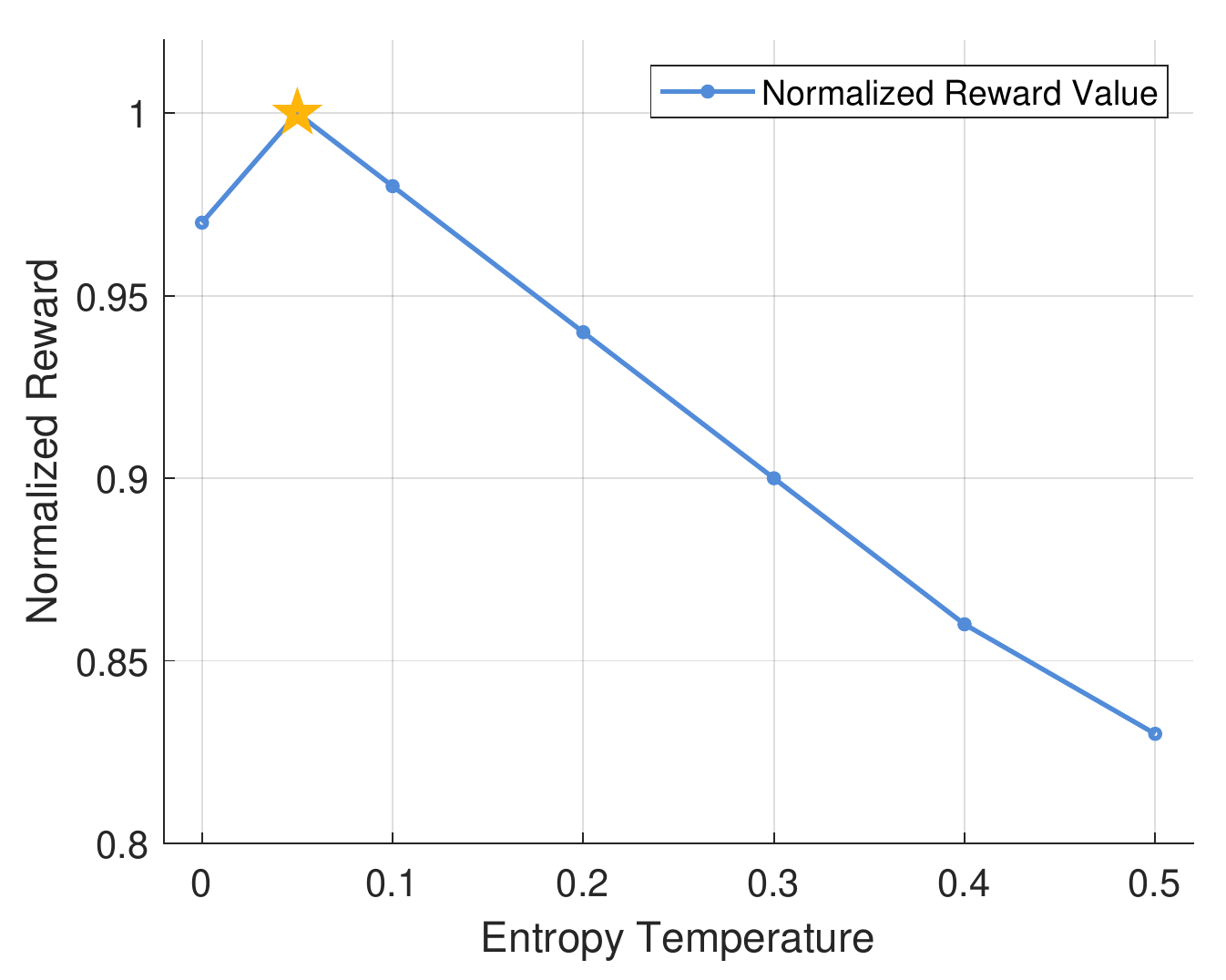}
			\vspace{-0.5cm}
			\caption{Effects of entropy regularization at different temperatures. Values are normalized by their maximum.}
			\label{fig:action-entropy}
		\end{minipage}
	\end{figure}

	
	\textbf{Understand the Reverse Diffusion Process.}
	Diffusion-based generative models employ the reverse diffusion process to generate new samples from a noise distribution. A denoising network is used to predict and remove noise at each step, gradually resulting in a high-quality and coherent sample. Fig.~\ref{fig:diffusion} illustrates how the distribution of action probability changes during each step of the reverse diffusion process at various training iterations. The starting point, represented by the step 0 column, is the softmax of a random normal distribution, which reflects the initial uncertainty of the diffusion model. As the process progresses, the decision probability, i.e., the output of the AGOD, becomes more peaked and approaches the optimal action predicted by the prophet policy, as shown by the vertical dotted lines.
	
	Figure \ref{fig:diffusion} highlights two important aspects of D2SAC. Firstly, as the learning process progresses, D2SAC can predict the optimal action decision probability distribution. This is evident in the third row of Fig.~\ref{fig:diffusion}, where D2SAC can successfully predict multiple optimal actions. Second, D2SAC maintains uncertainty after several {\color{black} denoising step} of denoising, allowing for exploration, which is crucial in DRL. However, as the number of {\color{black} denoising step} increases, the exploration ability decreases, leading to suboptimal solutions. This explains the reason for the decrease in reward in Fig.~\ref{fig:denoisingstep} when $T$ is larger than 5. The exploration-exploitation trade-off feature of D2SAC in discrete action spaces is distinct and novel, different from approaches in continuous action spaces. In the problem with continuous action spaces, other techniques, such as noise exploration, should be used to enhance exploration. Our approach is thus innovative and different from other approaches \cite{wang2022diffusion, janner2022planning}.
	
	\textbf{Balance Exploration and Exploitation with Action Entropy.}
	To balance exploration and exploitation in D2SAC, it is crucial to determine the strength of inherent exploration ability. A smaller value of the {\color{black} denoising step} $T$ can increase uncertainty, causing the agent to explore actions that may not yield high rewards. Conversely, a larger $T$ can decrease uncertainty but may cause the agent to stay with suboptimal solutions. The action entropy regularization proposed in \cite{haarnoja2018soft} addresses this challenge by adding a penalty to the expected reward, which is controlled by the temperature coefficient $\alpha$. This regularization balances the trade-off between exploration and exploitation by modulating the extent to which the agent can explore less likely actions.
	
	Figure \ref{fig:action-entropy} illustrates the impact of the action entropy regularization on the expected reward of D2SAC for varying entropy temperature values ($\alpha$). The results suggest an optimal value of $\alpha=0.05$, which balances exploration and exploitation performance. A lower $\alpha$ hinders the agent from selecting actions with high uncertainty, leading to greedy behavior and missing out on discovering better actions. Conversely, a higher $\alpha$ encourages the agent to become random, resulting in slow or no progress in learning the optimal policy. By maintaining an appropriate level of entropy, D2SAC achieves a balance between exploration and exploitation, resulting in a fast convergence to the optimal policy. 
	
	\section{Conclusion}\label{Scon}
	We have proposed an innovative {\color{black}edge-enabled} AaaS architecture to enable ubiquitous AIGC functionality. 
	To tackle the challenges of environmental uncertainty and variability, we have developed the AGOD based on the diffusion model, which is used in DRL to create the D2SAC algorithm for efficient and optimal ASP selection. 
	Our extensive experimental results have demonstrated the effectiveness of the proposed algorithm, which outperformed seven representative DRL algorithms in both the ASP selection problem and various standard control tasks. 
	Our proposed approach provides a practical and effective solution for the ubiquitous AIGC service in Metaverse. 
	{\color{black} More importantly}, the AGOD algorithm can potentially be applied to various optimization problems in wireless networks.
	{\color{black}In our future research, we intend to collect and employ real-world datasets related to edge-enabled ASP selections, allowing us to validate and refine our algorithm in practical scenarios effectively.}
	\bibliographystyle{IEEEtran} 
	{\small \bibliography{ref}}
\end{document}